\input ./style/arxiv-general.cfg
\documentclass[MSNbibl,nameyear,dvips]{arxstspdf}
\makeatletter
   \@ifpackageloaded{graphicx}{}{\usepackage{graphicx}}
\makeatother
\usepackage{flushend}
\usepackage{stfloats}

\volume{30}
\issue{2}
\pubyear{2015}
\firstpage{242}
\lastpage{257}
\doi{10.1214/14-STS510}
\docsubty{FLA}

\makeatletter
\def\cal{\mathcal}

\def\p{\mathrm{p}}
\newcommand{\eqref}[1]{(\ref{#1})}

\newcommand{\Nor}{\mathrm{N}} 
\newcommand{\Uni}{\mathrm{U}} 
\newcommand{\Ga}{\mathrm{G}} 
\newcommand{\IG}{\operatorname{IG}} 
\newcommand{\IncIG}{\operatorname{IncIG}} 
\newcommand{\E}{\mathbb{E}} 
\newcommand{\Mod}{\operatorname{Mod}} 
\newcommand{\diag}{\operatorname{diag}} 
\renewcommand{\P}{\operatorname{\mathsf{Pr}}} 

\newcommand{\sima}{\mathrel{\stackrel{a}{\thicksim}}} 

\newcommand{\given}{ \vert } 
\newcommand{\absmall}[1]{\vert #1\vert}
\newcommand{\IdMat}{\mathbf{I}} 

\newcommand{\fisher}{\mathcal{I}}
\newcommand{\mset}{\mathcal{J}}
\newcommand{\mi}{\mathcal{M}}
\newcommand{\TBFs}{\mathrm{TBF}}
\newcommand{\mTBFs}{\mathrm{mTBF}}
\newcommand{\TBF}[2]{\mathrm{TBF}_{#1,#2}}
\newcommand{\mTBF}[2]{\mathrm{mTBF}_{#1,#2}}
\newcommand{\DBF}[2]{\mathrm{DBF}_{#1,#2}}
\newcommand{\mDBF}[2]{\mathrm{mDBF}_{#1,#2}}
\makeatother

\begin{document}
\begin{frontmatter}

\title{Approximate Bayesian Model Selection with the Deviance Statistic}
\runtitle{Approximate Bayesian Model Selection with the Deviance}

\begin{aug}
\author[A]{\fnms{Leonhard}~\snm{Held}\corref{}\ead[label=e1]{leonhard.held@uzh.ch}},
\author[B]{\fnms{Daniel}~\snm{Saban\'es Bov\'e}\ead[label=e2]{daniel.sabanes\_bove@roche.com}}
\and
\author[A]{\fnms{Isaac}~\snm{Gravestock}\ead[label=e3]{isaac.gravestock@uzh.ch}}
\runauthor{L. Held, D. Saban\'es Bov\'e and I. Gravestock}

\affiliation{University of Zurich and F. Hoffmann-La Roche Ltd}

\address[A]{Leonhard Held is Professor and Isaac Gravestock is Ph.D.~Student,
Department of Biostatistics,
Institute of Epidemiology, Biostatistics and Prevention,
University of Zurich,
Hirschengraben 84, 8001 Zurich,
Switzerland \printead{e1,e3}.}
\address[B]{\mbox{Daniel~Saban\'es} Bov\'e is Biostatistician at F. Hoffmann-La
Roche Ltd, 4070 Basel, Switzerland \printead{e2}.}
\end{aug}

%
\begin{abstract}
Bayesian model selection poses two main challenges: the
specification of parameter priors for all models, and the
computation of the resulting Bayes factors between models. There is
now a large literature on automatic and objective parameter priors
in the linear model. One important class are \mbox{$g$-priors}, which were
recently extended from linear to generalized linear models
(GLMs). We show that the resulting Bayes factors can be approximated
by test-based Bayes factors (Johnson
[\textit{Scand. J. Stat.} \textbf{35}
(2008) 354--368]) using the deviance
statistics of the models. To estimate the hyperparameter $g$, we
propose empirical and fully Bayes approaches and link the former to
minimum Bayes factors and shrinkage estimates from the literature.
Furthermore, we describe how to approximate the corresponding
posterior distribution of the regression coefficients based on the
standard GLM output. We illustrate the approach with the
development of a clinical prediction model for 30-day survival in
the GUSTO-I trial using logistic regression.
\end{abstract}

%
\end{frontmatter}

\section{Introduction}
\label{sec:introduction}

The problem of model and variable selection is pervasive in
statistical practice. For example, it is central for the development
of clinical prediction models [\citet{Steyerberg2009}]. For
illustration, we consider the GUSTO-I trial, a large
randomized study for comparison of four different treatments in over
40,000 acute myocardial infarction patients [\citet{lee.etal1995}]. We
study a publicly available subgroup from the Western region of
the USA with $n=2188$ patients and prognosis of the binary endpoint
30-day survival [\citet{Steyerberg2009}]. In order to develop a
clinical prediction model for this endpoint, we focus our analysis on
the assessment of the effects of 17 covariates listed in
Table~\ref{tab:gusto-description} in a logistic regression model.
%
\begin{table*}
\caption{Description of the variables in the GUSTO-I data set}
\label{tab:gusto-description}
\begin{tabular*}{345pt}{@{\extracolsep{\fill}}ll@{}}
\hline
\textbf{Variable} & \textbf{Description} \\
\hline
$y$ & Death within 30 days after acute myocardial infarction
(Yes${} = 1$, No${} = 0$) \\
$x_{1}$ & Gender (Female${} = 1$, Male${} = 0$) \\
$x_{2}$ & Age [years] \\
$x_{3}$ & Killip class (4 categories) \\
$x_{4}$ & Diabetes (Yes${}= 1$, No${} = 0$) \\
$x_{5}$ & Hypotension (Yes${} = 1$, No${} = 0$) \\
$x_{6}$ & Tachycardia (Yes${} = 1$, No${} = 0$) \\
$x_{7}$ & Anterior infarct location (Yes${} = 1$, No${} = 0$) \\
$x_{8}$ & Previous myocardial infarction (Yes${}= 1$, No${}= 0$) \\
$x_{9}$ & Height [cm] \\
$x_{10}$ & Weight [kg] \\
$x_{11}$ & Hypertension history (Yes${}=1$, No${}=0$) \\
$x_{12}$ & Smoking (3 categories: Never/Ex/Current) \\
$x_{13}$ & Hypercholesterolaemia (Yes${}=1$, No${}=0$) \\
$x_{14}$ & Previous angina pectoris (Yes${}=1$, No${}=0$) \\
$x_{15}$ & Family history of myocardial infarctions (Yes${}=1$,
No${}=0$) \\
$x_{16}$ & ST elevation on ECG: Number of leads (0--11) \\
$x_{17}$ & Time to relief of chest pain more than 1 hour (Yes${}=1$,
No${}=0$) \\
\hline
\end{tabular*}
\end{table*}

There is now a large literature on automatic and objective Bayesian
model selection, which unburden the statistician from eliciting
manually the parameter priors for all models in the absence of
substantive prior information [see, e.g., \citet{BergerPericchi2001}].
However, such objective Bayesian methodology is currently limited to
the linear model [e.g., \citet{bayarri.etal2012}], where the $g$-prior
on the regression coefficients is the standard choice
[\citet{LiangPauloMolinaClydeBerger2008}]. For non-Gaussian regression,
there are computational and conceptual problems, and one solution to
this are test-based Bayes factors [\citet{johnson2005}]. Consider a
classical scenario with a null model nested within a more general
alternative model. Traditionally, the use of Bayes factors requires
the specification of proper prior distributions on all unknown model
parameters of the alternative model, which are not shared by the null
model. In contrast, \citet{johnson2005} defines Bayes factors using the
distribution of a suitable test statistic under the null and
alternative models, effectively replacing the data with the test
statistic. This approach eliminates the necessity to define prior
distributions on model parameters and leads to simple closed-form
expressions for $\chi^2$-, $F$-, $t$-, and $z$-statistics.

The \citet{johnson2005} approach is extended in \citet{johnson2008} to
the likelihood ratio test statistic and, thus, if applied to
generalized linear regression models (GLMs), to the deviance statistic
[\citet{NelderWedderburn1972}]. This is explored further in
\citet{hu.johnson2009}, where Markov chain Monte Carlo (MCMC) is used
to develop a Bayesian variable selection algorithm for logistic
regression. However, the factor $g$ in the implicit
$g$-prior is treated as fixed and estimation of the regression coefficients
is also not
discussed. We fill this gap and extend the work by
\citet{hu.johnson2009}, combining $g$-prior methodology for the linear
model with Bayesian model selection based on the deviance. This
enables us to apply empirical [\citet{GeorgeFoster2000}] and fully
Bayesian [\citet{CuiGeorge2008}] approaches for estimating the
hyperparameter $g$ to GLMs. By linking $g$-priors to the theory on
shrinkage estimates of regression coefficients
[\citeauthor{copas1983} (\citeyear{copas1983,Copas1997})], we finally obtain a unified framework
for objective Bayesian model selection and parameter inference for
GLMs.

The paper is structured as follows. In
Section~\ref{sec:merg-object-bayes} we review the $g$-prior in the
linear and generalized linear model, and show that this prior choice
is implicit in the application of test-based Bayes factors computed
from the deviance statistic. In Section~\ref{sec:calibrating-prior}
we describe how the hyperparameter $g$ influences model selection and
parameter inference, and introduce empirical and fully Bayesian
inference for it. Using empirical Bayes to estimate $g$, we are able
to analytically quantify the accuracy of test-based Bayes factors in
the linear model. Connections to the literature on minimum Bayes
factors and shrinkage of regression coefficients are outlined. In
Section~\ref{sec:applications} we apply the methodology in order to
build a logistic regression model for predicting 30-day survival in
the GUSTO-I trial, and compare our methodology with selected
alternatives in a bootstrap study. In Section~\ref{sec:discussion} we
summarize our findings and sketch possible extensions.

\section{Objective Bayesian Model Selection in Regression}
\label{sec:merg-object-bayes}

Consider a generic regression model $\mi$ with linear predictor $\eta
= \alpha+ \mathbf{x}^{\top}\bolds{\beta}$, from which we
assume that the outcome $\mathbf{y}=(y_{1},\ldots, y_{n})$ was
generated. We collect the intercept $\alpha$, the regression
coefficients vector~$\bolds{\beta}$, and possible additional
parameters (e.g., the residual variance in a linear model) in
$\bolds{\theta} \in\bolds{\Theta}$. Specific candidate
models $\mi_{j}$, $j \in\mset$, differ with respect to the content
and the dimension of the covariate vector $\mathbf{x}$, and hence
$\bolds{\beta}$, so each
model $\mi_{j}$ defines its own parameter vector $\bolds{\theta}_j$
with
likelihood function $\p(\mathbf{y}
\given\bolds{\theta}_{j}, \mi_{j})$.

Through optimizing this likelihood, we obtain the maximum
likelihood estimate (MLE) $\hat{\bolds{\theta}}_{j}$ of
$\bolds{\theta}_{j}$.
For Bayesian inference a prior distribution with density
$\p(\bolds{\theta}_{j} \given\mi_{j})$ is assigned to the
parameter vector
$\bolds{\theta}_{j}$ to obtain the posterior density $\p
(\bolds{\theta}_{j}
\given\mathbf{y}, \mi_{j}) \propto  \p(\mathbf{y} \given
\bolds{\theta}_{j}, \mi_{j}) \p(\bolds{\theta}_{j} \given
\mi_{j})$.
This forms the basis to compute the posterior mean
$\E(\bolds{\theta}_{j} \given\mathbf{y}, \mi_{j})$
and other suitable characteristics of the posterior distribution.
The marginal likelihood
\[
\p(\mathbf{y} \given\mi_{j}) = \int_{\bolds{\Theta}_{j}} \p(
\mathbf{y} \given\bolds{\theta}_{j}, \mi_{j}) \p(
\bolds{\theta}_{j} \given\mi_{j}) \,d\bolds{\theta}_{j}
\]
is the key ingredient to transform
prior model probabilities $\P(\mi_{j})$, $j \in\mset$, to posterior
model probabilities
%
\begin{eqnarray}
\label{eq:bf-to-post-prob} \P(\mi_{j} \given\mathbf{y}) &=& \frac
{\p(\mathbf{y} \given\mi_{j}) \P(\mi_{j})}{
\sum_{k \in\mset} \p(\mathbf{y} \given\mi_{k}) \P(\mi_{k})}
\nonumber
\\[-8pt]
\\[-8pt]
\nonumber
& =& \frac{\DBF{j}{0} \P(\mi_{j})}{\sum_{k \in\mset} \DBF{k}{0}
\P(\mi_{k})}.
\end{eqnarray}
In the second line, the usual (data-based) Bayes factor $\DBF{j}{0} =
\p(\mathbf{y} \given\mi_{j}) / \p(\mathbf{y} \given\mi_{0})$
of model $\mi_{j}$ {versus} a reference model $\mi_{0}$ replaces
the marginal likelihood $\p(\mathbf{y} \given\mi_{j})$ from
the first line. Improper priors can only be used
for parameters that are common to all models (e.g., here the intercept
$\alpha$), because only then the indeterminate normalizing constant
cancels in the posterior model probabilities
\eqref{eq:bf-to-post-prob}.

In Section~\ref{sec:g-priors} we discuss the $g$-prior, a specific
class of prior distributions $\p(\bolds{\theta}_{j} \given
\mi_{j})$, commonly used in linear model selection problems. The
$g$-prior induces shrinkage of $\bolds{\beta}$, in the sense that
the posterior mean is a shrunken version of the MLE toward the prior
mean. Furthermore, it is an automatic prior, since it does
not require specification of subjective prior information.
Section~\ref{sec:test-based-bayes} discusses the resulting test-based
Bayes factors under the $g$-prior.

\subsection{Zellner's $g$-Prior and Generalizations}
\label{sec:g-priors}

We start with the original formulation of Zellner's $g$-prior for the Gaussian
linear model in Section~\ref{sec:g-priors:gaussian-linear-model} and
extend this
to GLMs in Section~\ref{sec:g-priors:generalized-linear-model}.

\subsubsection{Gaussian linear model}
\label{sec:g-priors:gaussian-linear-model}

Consider the Gaussian linear model $\mi_{j}\dvtx y_{i} \sim
\Nor(\alpha+ \mathbf{x}_{ij}^{\top}\bolds{\beta}_{j},
\sigma^{2})$ with intercept $\alpha$, regression coefficients vector
$\bolds{\beta}_{j}$, and variance $\sigma^{2}$, and collect all
parameters in $\bolds{\theta}_{j} = (\alpha,
\bolds{\beta}_{j}^\top,  \sigma^{2})^\top$. Here $\Nor(\mu,
\sigma^{2})$ denotes the univariate Gaussian density with mean $\mu$
and variance $\sigma^{2}$, and $\mathbf{x}_{ij} = (x_{i1},\ldots,
x_{id_{j}})^{\top}$ is the covariate vector for observation $i=1,\ldots, n$.
Using the $n \times d_{j}$ full rank design matrix
$\mathbf{X}_{j} = (\mathbf{x}_{1j},\ldots,
\mathbf{x}_{nj})^{\top}$, the likelihood obtained from $n$
independent observations is
%
\begin{equation}
\label{eq:linear-model-mult-density} \p(\mathbf{y} \given\bolds{\theta}_{j},
\mi_{j}) = \Nor_{n} \bigl(\mathbf{y} \given\alpha
\mathbf{1}+ \mathbf{X}_{j}\bolds{\beta}_{j},
\sigma^{2} \IdMat \bigr),
\end{equation}
with $\mathbf{1}$ and $\IdMat$ denoting the all-ones vector and
identity matrix of dimension $n$, respectively. We assume that the covariates
have been centered around 0, that is,
$\mathbf{X}_{j}^{\top}\mathbf{1} = \mathbf{0}$. Here
and in the
following, $\mathbf{0}$ denotes the zero vector of length $d_j$.

Zellner's $g$-prior [\citet{Zellner1986}] fixes a constant $g>0$ and
specifies the Gaussian prior
%
\begin{equation}
\bolds{\beta}_{j} \given\sigma^{2}, \mi_{j}
\sim \Nor_{d_{j}} \bigl( \mathbf{0}, g \sigma^{2} \bigl(
\mathbf{X}_{j}^{\top} \mathbf{X}_{j}
\bigr)^{-1} \bigr) \label{eq:g-priors:gaussian-linear-model:g-prior}
\end{equation}
for the regression coefficients $\bolds{\beta}_{j}$, conditional
on $\sigma^2$. This prior can be interpreted
as a posterior distribution, if $\alpha$ is fixed and a locally uniform
prior for
$\bolds{\beta}_{j}$ is combined with an imaginary outcome
$\mathbf{y}_{0}= \alpha  \mathbf{1}$
from the Gaussian linear
model \eqref{eq:linear-model-mult-density} with the same design matrix
$\mathbf{X}_{j}$ but scaled residual variance $g\sigma^{2}$.
The prior \eqref{eq:g-priors:gaussian-linear-model:g-prior} on
$\bolds{\beta}_{j}$ is usually combined with
an improper reference prior on the intercept $\alpha$ and the residual
variance $\sigma^2$
[\citet{LiangPauloMolinaClydeBerger2008}]:
$\p(\alpha, \sigma^{2}) \propto\sigma^{-2}$.
The posterior distribution of $(\alpha,
\bolds{\beta}_{j}^{\top})^{\top}$ is then a multivariate $t$
distribution,
with posterior mean of $\bolds{\beta}_{j}$ given by
%
\begin{equation}
\label{eq:g-prior-post-mean-beta} \E(\bolds{\beta}_{j} \given\mathbf{y},
\mi_{j}) = \frac{g}{g+1} \hat{\bolds{\beta}}_{j} =
\frac{n\cdot\hat{\bolds{\beta}}_{j} + n/g\cdot\mathbf{0}}{n
+ n/g}.
\end{equation}
This means that the MLE $\hat{\bolds{\beta}}_{j}$, the ordinary
least squares (OLS) estimate,
is shrunk toward the prior mean zero.
The shrinkage factor $t=g/(g+1)$ scales the MLE to obtain the posterior
mean \eqref{eq:g-prior-post-mean-beta}. In other words, the
posterior mean is a weighted average of the MLE and the prior mean with
weights proportional to the data sample size $n$ and the term $n/g$,
respectively.
Thus, $n/g$ can be interpreted as the prior sample size, or $1/g$ as the
relative prior sample size. The question of how to choose or estimate
$g$ will be
addressed in Section~\ref{sec:calibrating-prior}.

One advantage of Zellner's $g$-prior is that the marginal likelihood, or,
equivalently, the (data-based) Bayes factor {versus} the null model
$\mi_{0}\dvtx {\bolds{\beta}}_{j} = \mathbf{0}$, has
a simple closed-form expression in terms of the usual coefficient of
determination
$R_{j}^{2}$ of model $\mi_{j}$ [\citet{LiangPauloMolinaClydeBerger2008}]:
%
\begin{eqnarray}
\label{eq:linear-model-g-prior-BF} &&\DBF{j} {0}
\nonumber
\\[-8pt]
\\[-8pt]
\nonumber
&&\quad= (g+1)^{(n-d_{j}-1)/2} \bigl\{1+g
\bigl(1-R_{j}^{2} \bigr) \bigr\}^{-(n-1)/2}.
\end{eqnarray}
Note that $R_{j}^{2}$ can be written as a function of the $F$-statistic
%
\begin{equation}
\label{eq:F-statistic} F_j = \bigl\{(n-d_{j}-1)
R_{j}^2 \bigr\}/ \bigl\{d_j
\bigl(1-R_{j}^2 \bigr) \bigr\}
\end{equation}
for testing ${\bolds{\beta}}_{j} = \mathbf{0}$. This
suggests that similar expressions (in terms of test statistics) can
be derived for the corresponding Bayes factors in GLMs. This
conjecture will be confirmed in Section~\ref{sec:test-based-bayes}.

\subsubsection{Generalized linear model}
\label{sec:g-priors:generalized-linear-model}

Now consider a GLM $\mi_{j}$ with linear predictor $\eta_{ij} =
\alpha
+ \mathbf{x}_{ij}^{\top}\bolds{\beta}_{j}$, mean $\mu
_{ij} =
h(\eta_{ij})$ obtained with the response function $h(\eta)$ and
variance function $v(\mu)$ [\citet{NelderWedderburn1972}]. The direct
extension of the standard $g$-prior in the Gaussian linear model is
then the generalized $g$-prior [\citet{sabanesbove.held2011}]
%
\begin{equation}
\bolds{\beta}_{j} \given\mi_{j} \sim
\Nor_{d_{j}} \bigl( \mathbf{0}, g c \bigl(\mathbf{X}_{j}^{\top}
\mathbf{W} \mathbf{X}_{j} \bigr)^{-1} \bigr),
\label{eq:g-priors:generalized-linear-model:g-prior}
\end{equation}
where $\mathbf{W}$ is a diagonal matrix with weights for the
observations (e.g., the binomial sample sizes for logistic
regression). Here the appropriate centering of the covariates is
$\mathbf{X}_{j}^{\top} \mathbf{W} \mathbf{1} =
\mathbf{0}$. As in
Section~\ref{sec:g-priors:gaussian-linear-model}, we specify an
improper uniform prior $\p(\alpha) \propto1$ for the intercept
$\alpha$. The constant $c = v\{h(\alpha)\} h'(\alpha)^{-2}$
[\citet{copas1983}; \citet{sabanesbove.held2011}]
in
\eqref{eq:g-priors:generalized-linear-model:g-prior}
corresponds to the variance $\sigma^{2}$ in the standard $g$-prior
\eqref{eq:g-priors:gaussian-linear-model:g-prior},
which could also be formulated for general linear models with a
nonunit weight matrix $\mathbf{W}$. It preserves the
interpretation of $n/g$ as the prior sample size. Note that
\citet{sabanesbove.held2011} recommend to use $\alpha=0$ as default,
but considerable improvements in accuracy can be obtained by using the
MLE $\hat\alpha$ of $\alpha$ under the null model;
see Section~\ref{sec:progn-modell-30:variable-selection} for details.

The connection between \eqref
{eq:g-priors:gaussian-linear-model:g-prior} and
\eqref{eq:g-priors:generalized-linear-model:g-prior} is as follows.
Denote the
expected Fisher information (conditional on the variance $\sigma^{2}$
in the
Gaussian linear model) for $(\alpha, \bolds{\beta}_{j}^{\top
})^{\top}$
as $\fisher(\alpha, \bolds{\beta}_{j})$. In the Gaussian
linear model,
this $(d_{j} + 1)\times(d_{j} + 1)$ matrix is block-diagonal due to the
centering of the covariates, and does not depend on the intercept nor the
regression coefficients:
\[
\fisher(\alpha, \bolds{\beta}_{j}) = %
\pmatrix{
\fisher_{\alpha, \alpha} & \fisher_{\alpha, \bolds{\beta
}_{j}} \vspace*{2pt}
\cr
\fisher_{\alpha, \bolds{\beta}_{j}}^{\top} & \fisher_{\bolds{\beta}_{j}, \bolds{\beta}_{j}} } %
=
\sigma^{-2} %
\pmatrix{ n & \mathbf{0}^{\top}
\vspace*{2pt}
\cr
\mathbf{0} & \mathbf{X}_{j}^{\top}
\mathbf{X}_{j} } %
.
\]
Hence, \eqref{eq:g-priors:gaussian-linear-model:g-prior} can be
written as
%
\begin{equation}
\label{eq:g-priors-generalized-linear-model:fisher-representation} \bolds{\beta}_{j} \given\mi_{j} \sim
\Nor_{d_{j}} \bigl( \mathbf{0}, g \cdot\fisher_{\bolds{\beta}_{j}, \bolds{\beta}_{j}}^{-1}
\bigr).
\end{equation}
In the GLM, $\fisher(\alpha, \bolds{\beta}_{j})$ depends on the
parameters and is not necessarily block-diagonal. However,
if we fix $\bolds{\beta}_{j}$ at its prior mean $\mathbf{0}$,
$\fisher(\alpha, \bolds{\beta}_{j} = \mathbf{0})$ is
block-diagonal with $\fisher_{\bolds{\beta}_{j},
\bolds{\beta}_{j}} = c^{-1}
\mathbf{X}_{j}^{\top}\mathbf{W} \mathbf{X}_{j}$, so
\eqref{eq:g-priors:generalized-linear-model:g-prior} and
\eqref{eq:g-priors-generalized-linear-model:fisher-representation} are
equivalent; see \citeauthor{copas1983} [(\citeyear{copas1983}), Section~8] for details. Departures
from the assumption $\bolds{\beta}_{j} = \mathbf{0}$ are also
discussed in \citet{copas1983}.

In contrast to Gaussian linear models, the marginal likelihood for GLMs
no longer has a closed-form expression. For its computation,
one has to resort to numerical approximations, for example, a Laplace
approximation. This requires a Gaussian approximation of the
posterior $\p(\alpha, \bolds{\beta}_{j} \given\mathbf{y},
\mi_{j})$, which can be obtained with the Bayesian iteratively
weighted least squares algorithm. See
\citeauthor{sabanesbove.held2011} [(\citeyear{sabanesbove.held2011}),
Section~3.1] for more details.

\subsection{Test-Based Bayes Factors}
\label{sec:test-based-bayes}

Based on the asymptotic distribution of the deviance statistic in
Section~\ref{SEC:TEST-BASED-BAYES:ASYMPT-DISTR-DEVI}, we connect the resulting
test-based Bayes factors with the $g$-prior in
Section~\ref{sec:test-based-bayes:defining-test-based} and discuss the
advantages over data-based Bayes factors in
Section~\ref{sec:test-based-bayes:advantages-tbf}.

\subsubsection{Asymptotic distributions of the deviance statistic}
\label{SEC:TEST-BASED-BAYES:ASYMPT-DISTR-DEVI}

Consider the frequentist approach to model selection, where test
statistics are used to assess the evidence against the null model
$\mi_{0}\dvtx \bolds{\beta}_{j} = \mathbf{0}$ in a specific GLM
$\mi_{j}$. A popular choice is the deviance
(or likelihood ratio test) statistic
\[
z_{j}(\mathbf{y}) = 2 \log \biggl\{ \frac
{\max_{\alpha, \bolds{\beta}_{j}} \p(\mathbf{y} \given
\alpha
, \bolds{\beta}_{j}, \mi_{j})}{
\max_{\alpha} \p(\mathbf{y} \given\alpha, \mi_{0})} \biggr\}.
\]
Then we have the well-known result that, conditional on $\mi_{0}$, the
distribution of the deviance $z_{j}(\mathbf{Y})$ converges for $n
\to\infty$ to a chi-squared distribution $\chi^{2}(d_{j})$ with
$d_{j}$ degrees of freedom.

To derive the asymptotic distribution of the deviance statistic under
model $\mi_{j}$, \citet{johnson2008} considers a sequence of local
alternative hypotheses $H_{1}^{n}\dvtx \bolds{\beta}_{j} =
\mathcal{O}(1 / \sqrt{n})$, so the size of the true regression
coefficients is scaled with $1/\sqrt{n}$, and thus gets smaller with
increasing number of observations $n$. This is the case of practical
interest, because for larger $\bolds{\beta}_{j}$ it would be
trivial to differentiate between $H_{0}\dvtx \bolds{\beta}_{j} =
\mathbf{0}$ and $H_{1}^{n}$, and for smaller
$\bolds{\beta}_{j}$ it would be too difficult
[\citet{johnson2005}, page 691]. In this setup, the distribution of the
deviance converges for $n \to\infty$ to a noncentral chi-squared
distribution $\chi^{2}(d_{j}, \lambda_{j})$ with $d_{j}$ degrees of
freedom, where $\lambda_{j} = \bolds{\beta}_{j}^{\top}
\fisher_{\bolds{\beta}_{j}, \bolds{\beta}_{j}}
\bolds{\beta}_{j}$ is the noncentrality parameter. Here
$\fisher_{\bolds{\beta}_{j}, \bolds{\beta}_{j}}$ denotes the
expected Fisher information for $\bolds{\beta}_{j}$ in model
$\mi_{j}$, evaluated at $\bolds{\beta}_{j} = \mathbf{0}$. See
Appendix \ref{app:proof-test-based-bayes} for a proof of this.

\subsubsection{Defining the test-based Bayes factor}
\label{sec:test-based-bayes:defining-test-based}
We now specify the generalized
$g$-prior \eqref
{eq:g-priors-generalized-linear-model:fisher-representation} for
$\bolds{\beta}_{j}$ in the alternative model $\mi_{j}$ with $g$
fixed. For the
noncentrality parameter $\lambda_{j} = \bolds{\beta}_{j}^{\top}
\fisher_{\bolds{\beta}_{j}, \bolds{\beta}_{j}}
\bolds{\beta}_{j}$, this corresponds to the gamma prior $\lambda
_{j} \sim
\Ga(d_{j}/2, 1/(2g))$ (see also
Appendix \ref{app:proof-test-based-bayes}). From above we have the approximate
``likelihood'' $z_{j} \given\lambda_{j} \sima\chi^{2}(d_{j},
\lambda
_{j})$ of the
deviance statistic $z_{j}$. \citet{johnson2008}, Theorem~2, shows that the
implied approximate marginal distribution of $z_{j}$
is
%
\begin{equation}
\label{eq:approx-marg-lik} z_{j} \sima\Ga \bigl(d_{j}/2, 1/ \bigl\{2
(g + 1) \bigr\} \bigr),
\end{equation}
which gives the approximate ``marginal likelihood''
$\p_{\mathrm{approx}}(z_{j} \given\mi_{j})$ of model $\mi_{j}$ in terms
of the
deviance statistic $z_{j}$. Furthermore, we have the approximate ``marginal
likelihood'' $\p_{\mathrm{approx}}(z_{j} \given\mi_{0})$ of the null model
$\mi_{0}$ from $z_{j} \sima\Ga(d_{j}/2, 1/2)$.
With these prerequisites, we can derive the \emph{test-based Bayes
factor} (TBF)
[\citet{johnson2008}]
%
\begin{eqnarray}\label{eq:tbf-definition}
\TBF{j} {0} &= &\frac{\p_{\mathrm{approx}}(z_{j} \given\mi_{j})}{
\p_{\mathrm{approx}}(z_{j} \given\mi_{0})}
\nonumber
\\[-8pt]
\\[-8pt]
\nonumber
&=& (g+1)^{-d_{j}/2} \exp \biggl(
\frac{g}{g+1} \frac{z_{j}}{2} \biggr)
\end{eqnarray}
of model $\mi_{j}$ {versus} model $\mi_{0}$ for fixed
$g$. $\TBF{j}{0}$ approximates the data-based Bayes factor $\DBF{j}{0}
= \p(\mathbf{y} \given\mi_{j}) / \p(\mathbf{y} \given
\mi_{0})$ obtained with the generalized
$g$-prior \eqref{eq:g-priors-generalized-linear-model:fisher-representation}.

It is instructive to compare the TBF \eqref{eq:tbf-definition} with
the DBF \eqref{eq:linear-model-g-prior-BF}
in the linear model if $g$ is fixed at the same value. Assume that
$0 < R_j^2 < 1$. Then we have $z_{j} = - n
\log(1 - R_{j}^{2})$ and
\eqref{eq:tbf-definition} can be written
as
$\TBF{j}{0}
= (g+1)^{-d_{j}/2}
(1 - R_{j}^{2})^{-   g   n /\{2(g+1)\}}$.
On the other hand, we have
\begin{eqnarray*}
\DBF{j} {0} &=& (g+1)^{(n-d_{j}-1)/2}\\
&&{}\cdot \bigl\{(g+1) \bigl(1-R_{j}^{2}
\bigr) + R_{j}^{2} \bigr\} ^{-(n-1)/2}
\\
& < & (g+1)^{(n-d_{j}-1)/2} \bigl\{(g+1) \bigl(1-R_{j}^{2}
\bigr) \bigr\}^{-(n-1)/2}
\\
& = & (g+1)^{-d_{j}/2} \bigl(1 - R_{j}^{2}
\bigr)^{- (n-1)/2}
\\
& = & \TBF{j} {0} \bigl(1 - R_{j}^{2} \bigr)^{\{1-n/(g+1)\}/2}
\\
& \leq& \TBF{j} {0}\quad \mbox{if } g \geq n-1.
\end{eqnarray*}
Hence, in the linear model, $\TBF{j}{0}$ will be larger than
$\DBF{j}{0}$ if both are calculated with the same $g \geq n-1$; however,
it is not clear which Bayes factor is larger for $g<n-1$.
In Section~\ref{sec:TBFs.vs.DBFs} we provide a
comparison of DBFs and TBFs in the case where $g$ is not fixed at the
same value, but estimated separately via empirical Bayes.
\subsubsection{Advantages of the test-based Bayes factor}
\label{sec:test-based-bayes:advantages-tbf}

Hu and Johnson (\citeyear{hu.johnson2009}) emphasize that TBFs behave like ordinary Bayes
factors, in the sense that for a sequence of nested models $\mi_{0}
\subset\mi_{1} \subset\mi_{2}$, we have $\TBF{2}{0} = \TBF{2}{1}
\cdot\TBF{1}{0}$. Hence, it is possible to compute coherent posterior
model probabilities from \eqref{eq:bf-to-post-prob} using TBFs in
place of DBFs. These probabilities will be invariant to the choice of
the baseline model $\mi_{0}$, in our case the null model. The
availability of posterior model probabilities is a clear advantage
over the $P$-values obtained from a classical analysis of deviance,
which are informal and indirect measures of evidence
[see, e.g., \citet{goodman19991}], and only suitable for pairwise model
comparisons. In addition, the Bayesian approach offers other posterior
probabilities of interest, for example, inclusion probabilities, which
are easy
to interpret and are required to compute the median probability model
[\citet{BarbieriBerger2004}].

Furthermore, the TBF can be computed much more easily than the DBF
because it only requires the deviance statistic $z_j$, which can by
calculated by standard GLM fitting software. No computation of the
expected Fisher information $\fisher_{\bolds{\beta}_{j},
\bolds{\beta}_{j}} = c^{-1}
\mathbf{X}_{j}^{\top}\mathbf{W} \mathbf{X}_{j}$ is
required, as it is only implicitly used in the prior formulation. In
contrast, the DBF does not have a closed form and thus needs to be
approximated by numerical means, which requires explicit calculation of the
inverse of $\fisher_{\bolds{\beta}_{j},
\bolds{\beta}_{j}}$. The computational advantages of TBFs over
DBFs increase further when $g$ is treated as unknown; see Section~\ref{sec:calibrating-prior}.

\section{Calibrating the {{${{g}}$}}-Prior}
\label{sec:calibrating-prior}

How does the prior variance factor $g$ in the generalized
$g$-prior \eqref{eq:g-priors-generalized-linear-model:fisher-representation}
influence posterior inference? We will look at the implications on shrinkage
and model selection in Section~\ref{sec:role-g-shrinkage}, and estimate
$g$ from
the data using empirical Bayes (Section~\ref{SEC:ESTIMATING-G-VIA-EB})
and fully
Bayes (Section~\ref{sec:full-bayes-estim}) procedures.

\subsection{The Role of $g$ for Shrinkage and Model Selection}
\label{sec:role-g-shrinkage}

We first look at the role of $g$ for shrinkage in a GLM, following the
arguments by \citet{copas1983}. It is well known from standard GLM
theory that the MLE $\hat{\bolds{\theta}}_{j} = (\hat{\alpha},
\hat{\bolds{\beta}}{}^{\top}_{j})^{\top}$ follows
asymptotically a normal distribution with mean
$\bolds{\theta}_{j}$ and covariance matrix equal to the inverse
expected Fisher information $\fisher(\alpha, \bolds{\beta}_{j})^{-1}$,
evaluated at the true values $\alpha$ and $\bolds{\beta}_{j}$.
As in \citet{copas1983}, we replace $\bolds{\beta}_{j}$
with its prior mean $\mathbf{0}$, that is, we assume that the
asymptotic inverse covariance matrix of $\hat{\bolds{\theta
}}_{j}$ is
$\fisher(\alpha, \mathbf{0}) =
\diag\{\fisher_{\alpha, \alpha}, \fisher_{\bolds{\beta}_{j},
\bolds{\beta}_{j}}\}$. Note that
$\hat{\alpha}$ and $\hat{\bolds{\beta}}_{j}$ are now uncorrelated
because we have centered the covariate vectors such that
$\mathbf{X}_{j}^{\top}\mathbf{W} \mathbf{1} =
\mathbf{0}$.

Combining this Gaussian ``likelihood'' of
$\bolds{\theta}_{j}$ with the generalized $g$-prior
\[
\bolds{\theta}_{j} \given g, \mi_{j} \sim
\Nor_{d_{j} + 1} \left( %
\pmatrix{ 0\vspace*{2pt}
\cr
\mathbf{0}
} %
, %
\pmatrix{ \infty& 0 \vspace*{2pt}
\cr
0 & g\cdot
\fisher_{\bolds{\beta}_{j}, \bolds{\beta}_{j}}^{-1} } %
 \right)
\]
gives the posterior distribution
%
\begin{eqnarray}\label{eq:approx-gaussian-posterior}
&&\bolds{\theta}_{j} \given\mathbf{y}, g, \mi_{j}
\nonumber
\\[-8pt]
\\[-8pt]
\nonumber
&&\quad\sim \Nor_{d_{j} + 1} \left( %
\pmatrix{ \hat{\alpha}\vspace*{2pt}
\cr
t \cdot\hat{\bolds{\beta}}_{j} } %
, %
\pmatrix{ \fisher_{\alpha, \alpha}^{-1} & 0 \vspace*{2pt}
\cr
0 & t \cdot
\fisher_{\bolds{\beta}_{j}, \bolds{\beta}_{j}}^{-1} } %
 \right).
\end{eqnarray}
Here $t = g / (g+1)$ is the same shrinkage factor for
$\hat{\bolds{\beta}}_{j}$ as in the Gaussian linear model from
Section~\ref{sec:g-priors:gaussian-linear-model}.
A smaller $g$ leads to a
smaller $t$ and thus to stronger shrinkage of the
$\bolds{\beta}_{j}$ posterior toward $\mathbf{0}$.
The approximate posterior
covariance matrix of $\bolds{\beta}_{j}$ is also shrunk by the
shrinkage factor $t$ compared to the frequentist covariance matrix.
In Section~\ref{sec:progn-modell-30:shrinkage} we provide an empirical
comparison of the true shrinkage under the generalized $g$-prior and
the theoretical shrinkage $g/(g+1)$.

The above assumption that the covariance matrix of the MLE is the
inverse expected Fisher information $\fisher(\alpha,
\mathbf{0})^{-1}$ enables us to derive a simple form of the
posterior distribution. In practice, we use the corresponding
sub-matrices of the observed Fisher information matrix evaluated at
the MLE, easily available from fitting a standard GLM, and
\eqref{eq:approx-gaussian-posterior} holds only approximately.
Likewise, the interpretion of $g$ as the ratio between the data sample
size and the prior sample size holds only approximately.

In order to understand the role of $g$ for model selection, consider
the TBF
formula \eqref{eq:tbf-definition} and the limiting case of $g \to0$.
Then the
generalized $g$-prior converges to a point mass at $\bolds{\beta
}_{j} =
\mathbf{0}$, and thus $\mi_{j}$ collapses to the null model
$\mi_{0}$. Consequently, $\TBF{j}{0} \to1$, because both models are equal
descriptions of the data in the limit.
On the other extreme, the case $g \to\infty$ corresponds to an increasingly
vague prior on $\bolds{\beta}_{j}$. As is well known,
arbitrarily inflating
the prior variance of parameters that are not common to all models is
not a safe
strategy. Here we see immediately from \eqref{eq:tbf-definition} that
$\TBF{j}{0} \to0$ in this case. This means that no matter how well
the model
$\mi_{j}$ fits the data compared to the null model $\mi_{0}$, the
latter is
preferred if $g$ is chosen large enough. This is an example of
Lindley's paradox
[\citet{Lindley1957}].

In between these two extremes, quite a few fixed values for $g$ have been
recommended.
The choice of $g=n$ corresponds to the unit information prior
[\citet{KassWasserman1995}], where the relative prior sample size is $1/n$.
For large $n$,
the TBF is asymptotically ($n \to\infty$) equivalent to the Bayesian
Information Criterion (BIC) [\citet{johnson2008}, page 358].
However, \citeauthor{hu.johnson2009} [(\citeyear{hu.johnson2009}), Section~3.1]
report that $g \in[2n, 6n]$
has led to favorable predictive properties and favorable operating
characteristics in a particular linear model variable selection example.
Other proposals in the linear model include the Risk Inflation
Criterion (RIC) by \citet{foster.george1994}, which sets
$g = d_{j}^{2}$, and the Benchmark prior by
\citet{FernandezLeySteel2001}, where $g = \max\{n, d_{j}^{2}\}$.

\subsection{Estimating $g$ via Empirical Bayes}
\label{SEC:ESTIMATING-G-VIA-EB}

The empirical Bayes (EB) approach [\citet{GeorgeFoster2000}] avoids
arbitrary choices of $g$ which may be at odds with the data. The local
EB approach, discussed in Section~\ref{sec:LEB}, retains computational
simplicity in comparison to the global EB approach, which we will
describe in Section~\ref{sec:GEB}. The local EB approach allows for an
analytic comparison of TBFs and DBFs in the linear model, as derived in
Section~\ref{sec:TBFs.vs.DBFs}.

\subsubsection{Local empirical Bayes}
\label{sec:LEB}
Consider one specific model $\mi_{j}$. If we choose $g$ such that
\eqref{eq:tbf-definition} is maximized, we obtain the estimate
%
\begin{equation}
\hat{g}_{\mathrm{LEB}} = \max\{z_{j}/d_{j} - 1, 0\}.
\label{eq:localEB-g-estimate}
\end{equation}
This is a local EB estimate because the prior parameter $g$ is
separately optimized in terms of the marginal likelihood
$\p_{\mathrm{approx}}(z_{j} \given\mi_{j})$ of each model $\mi_{j}$, $j
\in\mset$ [\citet{GeorgeFoster2000}]. Using these values of $g$, the
evidence in favor of the alternative hypothesis $H_1$ is maximized.
This has the disadvantage that the resulting maximum TBFs
%
\begin{eqnarray}
\label{eq:maximum-tbf} &&\mTBF{j} {0}
\nonumber
\\[-8pt]
\\[-8pt]
\nonumber
&&\quad= \max \biggl\{ \biggl( \frac{z_{j}}{d_{j}}
\biggr)^{-d_{j}/2} \exp \biggl( \frac{z_{j} - d_{j}}{2} \biggr), 1 \biggr\},
\end{eqnarray}
obtained by plugging \eqref{eq:localEB-g-estimate} into
\eqref{eq:tbf-definition}, are not consistent if the null model is
true [\citet{johnson2008}, page 355], that is, $\P(\mi_{0} \given y)
\not
\to1$
for $n\to\infty$ if $\mi_{0}$ is true. This is clear from above
because \eqref{eq:maximum-tbf} will always be larger than 1, instead
of converging to $0$, which is necessary for consistent accumulation
of evidence in favor of the null model.

However, the corresponding shrinkage factors
%
\begin{equation}
\label{eq:local-EB-shrinkage} \hat{t}_{\mathrm{LEB}} = \frac{\hat{g}_{\mathrm{LEB}}}{\hat
{g}_{\mathrm
{LEB}} + 1} = \max\{1 -
d_{j}/z_{j}, 0\}
\end{equation}
are exactly the same as proposed by \citeauthor{Copas1997}
[(\citeyear{Copas1997}), page 176] for
out-of-sample prediction. He developed this formula specifically for
logistic regression by generalizing the formula for linear models.
See also \citeauthor{vanhouwelingen.lecessie1990}
[(\citeyear{vanhouwelingen.lecessie1990}), page 1322] for another
justification
of this widely used shrinkage factor.

There is a close connection between maximum
TBFs \eqref{eq:maximum-tbf} and minimum Bayes factors, which are used
to transform $P$-values into lower bounds on the corresponding Bayes factor.
Just as TBFs, these methods
usually consider the value of a test statistic (or the corresponding
$P$-value) as the data
[\citet
{EdwardsLindmanSavage1963}; \citet{berger.sellke1987};
\citet{goodman19992}; \citet{sellke.etal2001}].
As already
noted by
\citet{held2010}, depending on the degrees of freedom $d_j$, the
maximum TBF \eqref{eq:maximum-tbf} turns out to be equivalent to
certain minimum Bayes factors (see Appendix \ref{app:proof-estimating-g-via-EB}
for explicit formulas and proofs):
For $d_{j}=1$, \eqref{eq:maximum-tbf} is equal to the \citet
{berger.sellke1987} bound for a
normal test statistic and a normal
prior on its mean.
For $d_{j}=2$, \eqref{eq:maximum-tbf} is equivalent to the \citet
{sellke.etal2001} bound.
For $d_{j} \rightarrow\infty$, \eqref{eq:maximum-tbf} is equal to the
\citet{EdwardsLindmanSavage1963} universal bound for one-sided $P$-values
obtained from normal test statistics.

The maximum TBF also has close connections to the Bayesian Local
Information Criterion (BLIC) proposed by
\citet{HjortClaeskens2003}, Section~9.2. The only difference is that
in the BLIC the deviance statistic is replaced by the squared Wald
statistic for testing $\bolds{\beta}_{j} =
\mathbf{0}$. However, the squared Wald statistic shares the same
noncentral chi-squared distribution as the deviance statistic in the
local asymptotic framework under the alternative model. Hence, the
BLIC could be considered as a possibly even more computationally
convenient approximation of the TBF in the sense of
\citet{lawless.singhal1978} who propose to replace the deviance
statistic with the squared Wald statistic for model selection
purposes. This comes at the price of losing the coherence of the TBF
for nested models described in
Section~\ref{sec:test-based-bayes:advantages-tbf}.

\subsubsection{Comparison with data-based Bayes factors}
\label{sec:TBFs.vs.DBFs}
We now continue the comparison of DBFs and TBFs in the
linear model from Section~\ref{sec:test-based-bayes:defining-test-based}, if the hyperparameter
$g$ is estimated with local empirical Bayes. For the DBFs
\eqref{eq:linear-model-g-prior-BF}, the local EB estimate of $g$ is
$\hat g = \max \{F_j - 1, 0  \}$, where $F_j$
is the $F$-statistic \eqref{eq:F-statistic};
see, for example, \citet{LiangPauloMolinaClydeBerger2008}, equation (9).
Plugging $\hat g$
into \eqref{eq:linear-model-g-prior-BF} gives
%
\begin{eqnarray}\label{eq:maximum-dbf}
&&\mDBF{j} {0}\nonumber\\
&&\quad = \max \bigl\{F_j^{(n-d_j-1)/2}
\bigl[F_j \bigl(1-R_j^2
\bigr)+R_j^2 \bigr]^{-(n-1)/2},1 \bigr\}
\nonumber
\\[-8pt]
\\[-8pt]
\nonumber
&& \quad = \max \biggl\{ \biggl(\frac{(n-1) R_j^2}{d_j} \biggr)^{-d_j/2} \\
&&\hspace*{22pt}\qquad{}\cdot\biggl(
\frac{1-R_j^2}{1-{d_j}/({n-1})} \biggr)^{-(n-d_j-1)/2} ,1 \biggr\} .\nonumber
\end{eqnarray}

A comparison of \eqref{eq:maximum-dbf} with \eqref{eq:maximum-tbf}
allows us to quantify the accuracy of mTBFs in the
Gaussian linear model.
First note that $1-R_j^2 = \exp(-z_j/n)$, so $R_j^2/(1-R_j^2) = \exp
(z_j/n)-1$.
Hence, $F_j \leq1$ if $z_j \leq d_j$, that is, $\mDBF{j}{0}=1$ if
$\mTBF{j}{0}=1$, and the error $\Delta= \log\mTBF{j}{0} - \log
\mDBF{j}{0}$
is nonnegative, if $\mDBF{j}{0}=1$. For $\mDBF{j}{0}>1$,
the second-order
Taylor approximation $R_j^2 \approx1-\exp(-z_j/n) \approx z_j/n
  \{1-z_j/(2n)\}$ in the first term of
\eqref{eq:maximum-dbf}
gives
%
\begin{eqnarray} \label{eq:maximum-tbf-approx}
&&\log\mDBF{j} {0}\nonumber\\
&&\quad \approx -\frac{d_j}{2} \biggl[\log(n-1) + \log \biggl(
\frac{z_j}{d_j} \biggr)+ \log \biggl(1-\frac{z_j}{2n} \biggr) \nonumber\\
&&\hspace*{182pt}{}- \log(n)
\biggr]
\\
&&\qquad{}+\frac{n-d_j-1}{2} \biggl(\frac{z_j}{n} - \frac{d_j}{n-1} \biggr)
\nonumber\\
&&\quad \approx -\frac{d_j}{2}\log \biggl(\frac{z_j}{d_j} \biggr) +
\frac{d_j   z_j}{4n} +\frac{n-d_j-1}{n} \cdot\frac{z_j-d_j}{2},\nonumber
\end{eqnarray}
where we have used the first-order
approximation $\log(1-x) \approx
-x$ both for $x=d_j/(n-1)$ and for $x=z_j/(2n)$
and have replaced $n-1$ with $n$, where suitable.

Comparing equation
\eqref{eq:maximum-tbf-approx} with \eqref{eq:maximum-tbf} 
finally reveals
that the error $\Delta$
is approximately
%
\begin{equation}
\label{eq:Delta} \widetilde{\Delta}  =  \max \biggl\{ \frac{d_j+1}{2n}
({z_j} - d_j) - \frac{d_j   z_j}{4n}, 0 \biggr\}.
\end{equation}
This is an interesting result. First,
$\widetilde{\Delta}$ is
positive so the mTBFs will tend to be larger than the corresponding
mDBFs. Second, the error is approximately linear in the deviance
$z_j$ and inversely related to the sample size~$n$. However,\vspace*{1pt} for fixed
$R_j^2$ the deviance $z_j$ grows linearly with $n$, which shows that
the error $\Delta$ is approximately independent of the sample size.
Finally, this formula suggests a simple bias-correction of mTBFs in
GLMs by multiplying \eqref{eq:maximum-tbf} with
$\exp(-\widetilde{\Delta})$, which we will apply in Section~\ref{sec:progn-modell-30:variable-selection}. We note that the
approximation \eqref{eq:Delta} is fairly accurate as long as $z_j/n$
is not too large, say, $z_j/n < 1$.
\subsubsection{Global empirical Bayes}
\label{sec:GEB}
An alternative EB approach is to maximize the weighted sum of the
TBFs with weights equal to the prior model probabilities, that is, to maximize
%
\begin{equation}
\sum_{j \in\mset} \TBF{j} {0} \Pr(\mi_{j})
\label{eq:global-marginal-likelihood}
\end{equation}
with respect to $g$. The resulting estimate $\hat{g}_{\mathrm{GEB}}$
parallels the global EB estimate
[\citet{LiangPauloMolinaClydeBerger2008}, Section~2.4] based on DBFs and\vadjust{\goodbreak}
needs to be computed by numerical optimization
of \eqref{eq:global-marginal-likelihood}. It was investigated by
\citet{GeorgeFoster2000} for the Gaussian linear model. Calculating
$\hat{g}_{\mathrm{GEB}}$ is more costly than calculating the
model-specific $\hat{g}_{\mathrm{LEB}}$, and is even infeasible when
$\absmall{\mset}$ is very large. In this case one could first perform
a stochastic model search
and then restrict the sum
in \eqref{eq:global-marginal-likelihood} to the set $\hat{\mset}$ of
models visited. The stochastic model search could be based on the
local EB estimates, say, and the resulting posterior model
probabilities are then ``corrected'' using the global EB estimate.

\subsection{Full Bayes Estimation of $g$}
\label{sec:full-bayes-estim}

EB approaches ignore the uncertainty of the
estimates $\hat{g}_{\mathrm{LEB}}$ and $\hat{g}_{\mathrm{GEB}}$,
respectively. As an alternative, we will now discuss fully Bayesian
estimation of $g$ using a continuous hyperprior for $g$. Thus, we
obtain continuous
mixtures of generalized $g$-priors, which we call generalized
hyper-$g$ priors [\citet{sabanesbove.held2011}]. Mixtures of $g$-priors
for model selection in the linear model were studied by
\citet{LiangPauloMolinaClydeBerger2008}.

\subsubsection{Priors for $g$}
\label{sec:full-bayes-estim:conjugate-prior}

In order to retain a closed form for the marginal likelihood of the model
$\mi_{j}$, the prior for $g$ must be conjugate to the (approximate)
``likelihood''
\[
\p_{\mathrm{approx}}(z_{j} \given g, \mi_{j}) \propto (g +
1)^{-d_{j}/2} \exp \biggl( - \frac{z_{j}/2}{g + 1} \biggr),
\]
obtained from \eqref{eq:approx-marg-lik}.
From this we see that an inverse-gamma prior $\IG(a, b)$ on $g + 1$, truncated
appropriately to the range $(1, \infty)$, is conjugate
[\citet{CuiGeorge2008}, page 891]. The corresponding prior density
function on $g$
is
%
\begin{equation}
\label{eq:inc-inv-gamma-prior-density} \quad\p(g) = \mathrm{M}(a, b) (g + 1)^{-(a + 1)} \exp \biggl(-
\frac{b}{g +
1} \biggr),
\end{equation}
where $\mathrm{M}(a, b) = b^{a}\{\int_{0}^{b}u^{a - 1}\exp(-u) \,du\}
^{-1}$ is the
normalizing constant. We denote this incomplete inverse-gamma distribution
as $g \sim\IncIG(a, b)$.
The model-specific posterior density then is
%
\begin{equation}
\label{eq:marg.post.g} g \given z_{j}, \mi_{j} \sim\IncIG(a +
d_{j}/2, b + z_{j}/2).
\end{equation}
Hence, the marginal likelihood of model $\mi_{j}$ is
\begin{eqnarray*}
\p(z_{j} \given\mi_{j}) &=& \frac{\p_{\mathrm{approx}}(z_{j} \given g, \mi_{j})\p(g)}{
\p(g \given z_{j}, \mi_{j})}
\\
&=& \frac{\mathrm{M}(a, b) z_{j}^{d_{j}/2 - 1}}{
\mathrm{M}(a + d_{j}/2, b + z_{j}/2) 2^{d_{j}/2} \Gamma(d_{j}/2)},
\end{eqnarray*}
and dividing this with $\p_{\mathrm{approx}}(z_{j} \given\mi_{0})$
finally yields
\[
\TBF{j} {0} = \frac
{\mathrm{M}(a, b)}{
\mathrm{M}(a + d_{j}/2, b + z_{j}/2)} \exp(z_{j}/2).
\]
A useful analytic consequence of \eqref{eq:marg.post.g} is that the
mode of the shrinkage factor
$t$ is
%
\begin{equation}\qquad
\Mod(t \given z_{j}, \mi_{j}) = \max \biggl\{1 -
\frac{a + d_{j}/2 - 1}{b + z_{j}/2}, 0 \biggr\}. \label{eq:inc-ig-posterior-mode}
\end{equation}

If the prior for $g$ is not conjugate, the required integration
of \eqref{eq:approx-marg-lik}, $\p(z_{j} \given\mi_{j}) = \int
\p_{\mathrm{approx}}(z_{j} \given g, \mi_{j}) \cdot\break \p(g)  \,dg$, can be
performed by one-dimensional numerical integration. Two examples of
nonconjugate hyperpriors on $g$ which are used in the Gaussian linear
model are the \citet{ZellnerSiow1980} prior, where $g \sim\IG(1/2,
n/2)$, and the hyper-$g/n$ prior proposed by
\citet{LiangPauloMolinaClydeBerger2008}:
%
\begin{equation}
\label{eq:hyper-g/n-prior} \frac{g/n}{g/n + 1} \sim\Uni(0, 1).
\end{equation}
Both priors give considerable probability mass to $g$ values
proportional to
$n$: The mode for the Zellner--Siow prior is $n/3$, and the median for the
hyper-$g/n$ prior is~$n$.

\subsubsection{Choice of hyperparameters}
\label{sec:full-bayes-estim:choos-hyperp}

The next question is then how to choose the hyperparameters $a, b$ of the
conjugate prior \eqref{eq:inc-inv-gamma-prior-density}.
\citet{CuiGeorge2008} recommend $a = 1$ and $b = 0$, which leads to
%
\begin{equation}
t = \frac{g}{g+1} \sim\Uni(0, 1), \label{eq:uniform-prior-on-shrinkage}
\end{equation}
a uniform prior on the shrinkage factor $t$. This is the hyper-$g$
prior by \citet{LiangPauloMolinaClydeBerger2008}, a~proper prior with
normalizing constant defined as the limit $\lim_{b \to0}
\mathrm{M}(a, b) = a$. The model-specific posterior
mode \eqref{eq:inc-ig-posterior-mode} of $t$ now equals the local EB
estimate $\hat{t}_{\mathrm{LEB}}$ in~\eqref{eq:local-EB-shrinkage},
as it
should, since we have used the uniform prior
\eqref{eq:uniform-prior-on-shrinkage} on $t$. Moreover, the marginal
posterior mode of~$t$, taking into account all models, will equal the
global EB estimate $\hat{t}_{\mathrm{GEB}} =
\hat{g}_{\mathrm{GEB}}/(\hat{g}_{\mathrm{GEB}}+1) $. This indicates that
using a hyper-$g$ prior will lead to similar results as the EB
methods.
Alternatively, matching the mode $n/3$ of the Zellner--Siow (ZS) prior
$g \sim\IG(1/2, n/2)$ suggests to use $g \sim\IncIG(a=1/2, b=(n +
3)/2)$. We call this the ZS adapted prior. The
posterior mode of $t$ is now $\Mod(t \given z_{j}, \mi_{j}) = 1 -
(d_{j} - 1)/(z_{j} + n + 3)$, which is always larger than
$\hat{t}_{\mathrm{LEB}}$ in \eqref{eq:local-EB-shrinkage} and thus
leads to weaker shrinkage of the regression coefficients.

The ZS prior and our adaptation depends on the sample size $n$, which
leads to consistent model selection, even if the null model is
true. Indeed, \citet{johnson2008} shows that for $g = \mathcal{O}(n)$
the TBF is consistent, because then the covariance matrix of the
generalized $g$-prior
\eqref{eq:g-priors:generalized-linear-model:g-prior} is
$\mathcal{O}(1)$ and prevents the alternative model from collapsing
with the null model. Here we have prior mode $n/3$, which fulfils
this condition. By contrast, the hyper-$g$
prior \eqref{eq:uniform-prior-on-shrinkage} has its median at $1$,
which clearly does not fulfil the condition. Moreover, the
model-specific posterior mode under the hyper-$g$ prior equals the
local EB estimate, which is inconsistent if the null model is true;
see Section~\ref{SEC:ESTIMATING-G-VIA-EB}. The hyper-$g/n$
prior \eqref{eq:hyper-g/n-prior} corrects this by scaling the prior to
have median $n$. However, these priors lead to weaker shrinkage than
the local EB approach or the hyper-$g$ prior. Stronger shrinkage as
in the empirical Bayes approaches is in general advantageous for
prediction [\citeauthor{copas1983} (\citeyear{copas1983,Copas1997})].

\subsubsection{Posterior parameter estimation}
\label{sec:full-bayes-estim:post-param-estim}

For a given GLM $\mi_{j}$ with deviance statistic $z_{j}$, we would
like to estimate the posterior distribution of its parameters
$\bolds{\theta}_{j} = (\alpha, \bolds{\beta}_{j}^\top
)^\top$.
We do
this by sampling from an approximation of the posterior distribution
\[
\p(\bolds{\theta}_{j} \given\mathbf{y}, \mi_{j}) =
\int \p(\bolds{\theta}_{j} \given g, \mathbf{y},
\mi_{j}) \p(g \given\mathbf{y}, \mi_{j}) \,dg,
\]
where we replace the data-based posterior $\p(g \given\mathbf{y},
\mi_{j})$ with the test-based posterior $\p(g \given z_j, \mi_{j})$ to
retain computational simplicity.

If a conjugate incomplete inverse-gamma prior distribution is
specified for $g$, we first need to sample from its model-specific
(test-based) posterior \eqref{eq:marg.post.g}. Sampling from an
$\IncIG
(a, b)$
distribution \eqref{eq:inc-inv-gamma-prior-density} is easy using
inverse sampling via its quantile function
\begin{eqnarray*}
&&\mathrm{F}^{-1}_{\IncIG(a, b)}(x) \\
&&\quad= %
\cases{\displaystyle
\frac
{b}{
\mathrm{F}^{-1}_{\IG(a, 1)}
\{(1 - x) \mathrm{F}_{\IG(a, 1)}(b)\}} - 1, & $b > 0$,\vspace*{2pt}
\cr
(1 - x)^{-1/a} - 1, &
$b = 0$, } %
\end{eqnarray*}
which is given in terms of the quantile and cumulative distribution
functions of the $\IG(a, 1)$ distribution. If a nonconjugate prior is
specified for $g$, then numerical methods can be used to sample from
$\p(g \given z_j, \mi_{j})$. Specifically, we approximate the log
posterior density using a linear interpolation, which is a by-product of
the numerical integration to obtain the marginal likelihood of the
model $\mi_{j}$.

In the second step, we sample the actual model parameters
$\bolds{\theta}_{j}$ from their approximate
posterior \eqref{eq:approx-gaussian-posterior} given the sample for
$g$. We use the observed Fisher information matrix, invert the
corresponding sub-matrices for $\hat{\alpha}$ and
$\hat{\bolds{\beta}}_{j}$, and scale the latter one with $t = g /
(g + 1)$. The MLE
$\hat{\bolds{\beta}}_{j}$ is also multiplied with $t$ to obtain the
appropriate mean of
the conditional Gaussian distribution \eqref{eq:approx-gaussian-posterior}.

\section{Application}
\label{sec:applications}

We consider data on 30-day survival from the GUSTO-I trial data as
introduced in Section~\ref{sec:introduction} and use the TBF
methodology as implemented in the \texttt{R}-package
``\texttt{glmBfp}'' available from \texttt{R}-Forge.\footnote{To
install the \texttt{R}-package, just type
\texttt{install.packages} (\texttt{"glmBfp", repos=}"
\surl{http://r-forge.r-project.org}")
into \texttt{R}.}

\subsection{Variable Selection}
\label{sec:progn-modell-30:variable-selection}

As there are $17$ explanatory variables in this data set, there are
$\absmall{\mset} = 2^{17} = 131\mbox{,}072$ different models to be
considered for variable selection. This is still a manageable size and
we can evaluate all models easily with TBFs (relative to the null
model) within a few minutes. In the absence of subjective prior
information on the importance of covariates, we use prior
inclusion probabilities of $1/2$ for each covariate and a marginal
uniform prior on
$d_{j}$. This is a commonly used objective prior assumption
[\citet{geisser1984}; \citet{scott.berger2010}].

We consider 4 approaches to estimate $g$:
local EB, the hyper-$g$ prior, the hyper-$g/n$ prior, and the ZS
adapted prior. Numerical computation of the corresponding DBFs
[\citet{sabanesbove.held2011}] is---depending on the method to estimate
$g$---between 11 (local EB) and 50 (ZS adapted prior) times
slower and requires explicit specification of the $g$-prior
\eqref{eq:g-priors:generalized-linear-model:g-prior}, including the
constant $c = v\{h(\alpha)\} h'(\alpha)^{-2}$. As $\alpha$ is unknown,
we fix it at the MLE $\hat\alpha$ obtained from the null model. We
will use this example to quantify the accuracy of the approximation of
DBFs by TBFs.
%
\begin{figure*}

\includegraphics{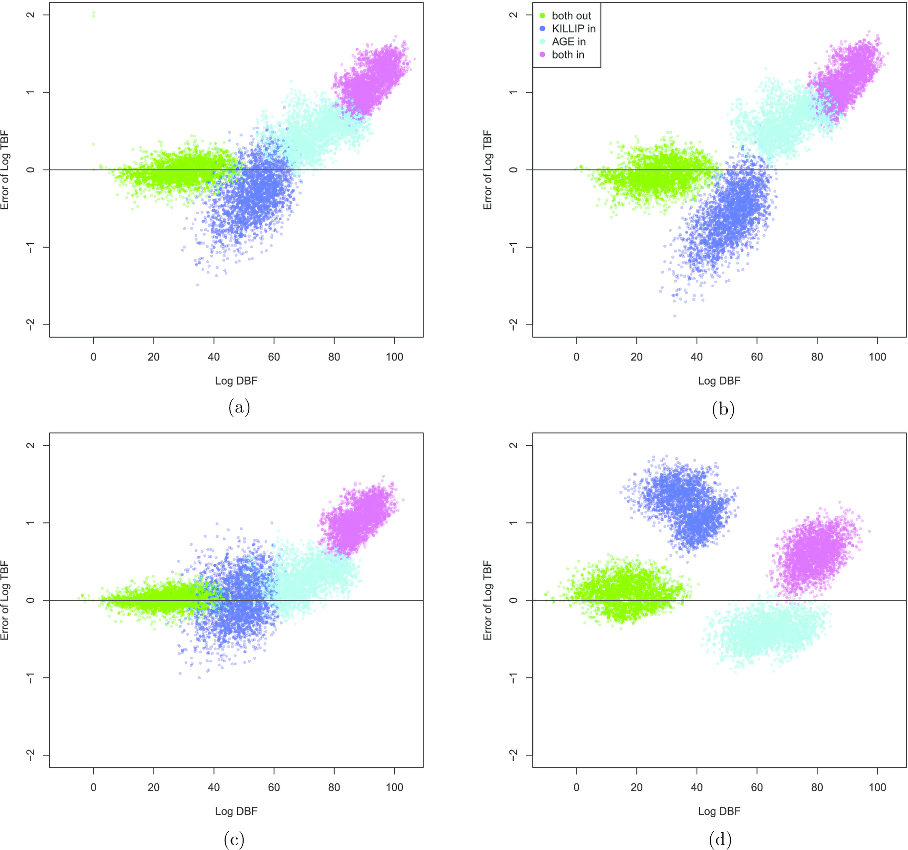}

\caption{Comparing test-based (TBF) and data-based (DBF)
log Bayes factors. The Bayes factors are shown in four different colors,
depending on whether or not the explanatory variables $x_2$ (Age)
and $x_3$ (Killip class) are included in the corresponding models.
\textup{(a)}~Local EB. \textup{(b)} Hyper-$g$. \textup{(c)}
Hyper-$g/n$. \textup{(d)} ZS adapted.}
\label{fig:gusto:compare-logBFs}
\end{figure*}

\begin{figure*}[b]

\includegraphics{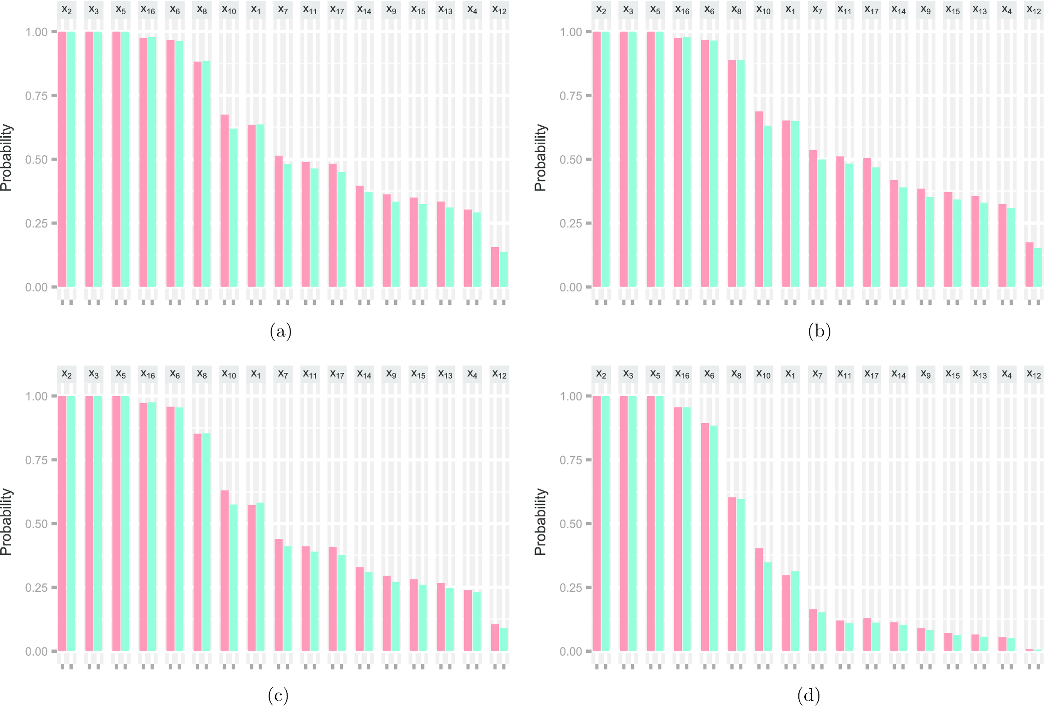}

\caption{Inclusion probabilities for all approaches,
comparing the
data-based (left bars, \protect
\includegraphics{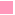}
) and the test-based
approach (right bars, \protect
\includegraphics{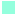}
). The
covariates are
ordered with respect to the results from the data-based approach under the
hyper-$g/n$ prior. \textup{(a)} Local EB. \textup{(b)} Hyper-$g$.
\textup{(c)} Hyper-$g/n$. \textup{(d)} ZS adapted.}
\label{fig:gusto:compare-var-select-inc-probs}
\end{figure*}

In Figure~\ref{fig:gusto:compare-logBFs}, we plot the error log TBF
$-$ log DBF against log DBF using the 4 different methods to estimate
$g$. To reduce the size of the figures, we
only show a random sample of 10,000 Bayes factors.
We note that the log DBFs vary between 0 and 106.7 (for local EB, where
the log Bayes factors cannot be negative),
$-0.7$ and 103.5 (\mbox{hyper-$g$}), $-6.8$ and 102.9 (hyper-$g/n$), and
$-14.1$ and
97.3 (under the ZS adapted prior).
On average, the log TBFs tend to be slightly larger than the log DBFs
with mean difference between 0.28 (hyper-$g$) and 0.37 (ZS
adapted). The standard deviations of the errors vary between 0.47
(hyper-$g/n$) and 0.70 (\mbox{hyper-$g$}). All Bayes factors for all four
methods had absolute error less than 2, apart from 12 TBFs calculated
with the EB approach, where the log DBF was zero, but the log TBF was
larger than zero.

\begin{figure*}

\includegraphics{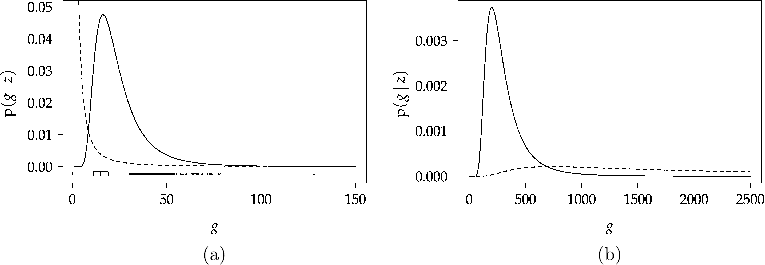}

\caption{Comparison of priors (dashed lines) and posteriors
(solid lines) of $g$ under the conjugate incomplete inverse-gamma
prior with
hyper-$g$ (left) and ZS adapted (right) hyperparameter
choices. \textup{(a)} Hyper-$g$ prior and posterior, together with
local EB
(boxplot for the values at bottom of the plot) and global EB (vertical line)
estimates of $g$. \textup{(b)} ZS adapted prior and posterior, together
with $g=n$ (vertical line).}
\label{fig:gusto:posterior-on-g}
\end{figure*}

Closer inspection of Figure~\ref{fig:gusto:compare-logBFs} reveals
that the error under the hyper-$g$ prior has a pattern similar to
that under the local EB approach.
For log DBFs larger than~50, the error of the TBFs tends to
increase with increasing DBFs, a feature that is visible in all 4
figures and to be expected from the approximate error \eqref{eq:Delta}
in the linear model.
Note that there is strong clustering visible for all four approaches
depending on whether or not the two most important explanatory
variables, $x_2$ (Age) and $x_3$ (Killip class), are included. The
corresponding four groups are given in different colors in
Figure~\ref{fig:gusto:compare-logBFs}. If both are included, the log
DBFs are large and the error of the TBFs is nearly always positive, a
feature that is present in all four approaches. Likewise, if the two
variables are not included, the Bayes factors are small and the
absolute error is close to zero. If one of the two is included, then
the size and direction of the error depends on the approach used.
Clustering is particularly pronounced for the ZS adapted prior,
where---somewhat surprisingly---the error of the log TBFs with $x_2$
excluded and $x_3$ included is around 1, whereas the error of the log TBFs
with $x_2$ included and $x_3$ excluded is negative, although the
corresponding DBFs tend to be larger. Thus, in this case the error
does not seem to increase in a monotone fashion with the DBFs.

Following the good agreement of TBFs and DBFs, the corresponding
posterior variable inclusion probabilities
are also very similar; see
Figure~\ref{fig:gusto:compare-var-select-inc-probs}. The two
neighboring bars have almost the same height for all covariates and
in all settings. The only exception is the variable Weight
($x_{10}$), where the difference is between 5 and 6 percentage points.
However, there are substantial differences in the inclusion
probabilities obtained with the different methods to estimate $g$. As
in the linear model [\citet{LiangPauloMolinaClydeBerger2008}], the ZS
adapted prior, favoring large values of $g$, leads to more
parsimonious models than the other three approaches. For example, the
local EB median probability model (MPM) under the TBF approach includes
the eight variables $x_1$, $x_2$, $x_3$, $x_5$, $x_6$, $x_8$,
$x_{10}$, $x_{16}$. Exactly the same model is selected under the
hyper-$g$ and the hyper-$g/n$ prior, whereas the MPM model under the
ZS adapted prior drops the variables $x_1$ and $x_{10}$ and includes
only the remaining six variables.

In Figure~\ref{fig:gusto:posterior-on-g}, the posterior distributions
of $g$ are compared with the underlying conjugate prior distributions
(ZS adapted and hyper-$g$) and local as well as global EB
estimates of $g$. The posterior distributions are based on all models
and computed using the identity
\[
\p(g \given\mathbf{z}) = \sum_{j\in\cal{J} }\p(g \given
z_j, \mi_j) \Pr (\mi_j \given
z_j).
\]
We clearly see the difference between the two priors resulting from
the different hyperparameter choices. The fixed choices $g=n$ (BIC)
and $g=2  n$ are not supported by the data, as all estimates are far
below these values. The local EB estimates of $g$ tend to be small, with
the posterior mode of $g$ under the hyper-$g$ prior and the global
EB estimate having similar values. The posterior mode of $g$ under the
ZS adapted prior is larger than the
other estimates but still much smaller than the fixed choices.

\subsection{Shrinkage of Coefficients}
\label{sec:progn-modell-30:shrinkage}
We now consider the MPM model identified in the previous section with
either the local EB, hyper-$g$, or hyper-$g/n$ approach, which includes
the eight variables $x_{1}$, $x_{2}$, $x_{3}$, $x_{5}$, $x_{6}$,
$x_{8}$, $x_{10}$, and $x_{16}$.
Integrated nested Laplace approximations [Rue, Martino and Chopin (\citeyear{naujas})]
have
been used to fit Bayesian logistic regression models under the
generalized $g$-prior for various values of $g$ with the R-INLA
package (\surl{www.r-inla.org}).
The constant $c$ in
\eqref{eq:g-priors:generalized-linear-model:g-prior} has been fixed
based on the estimate $\hat\alpha$ of $\alpha$ in the null model.
Empirical shrinkage is defined as the ratio of the resulting posterior
mean estimates of the regression coefficients over the corresponding
MLEs. Empirical shrinkage can also be computed based on the ratio of
the resulting posterior variances over the corresponding variances of
the MLEs; compare equation \eqref{eq:approx-gaussian-posterior}.

\begin{figure*}

\includegraphics{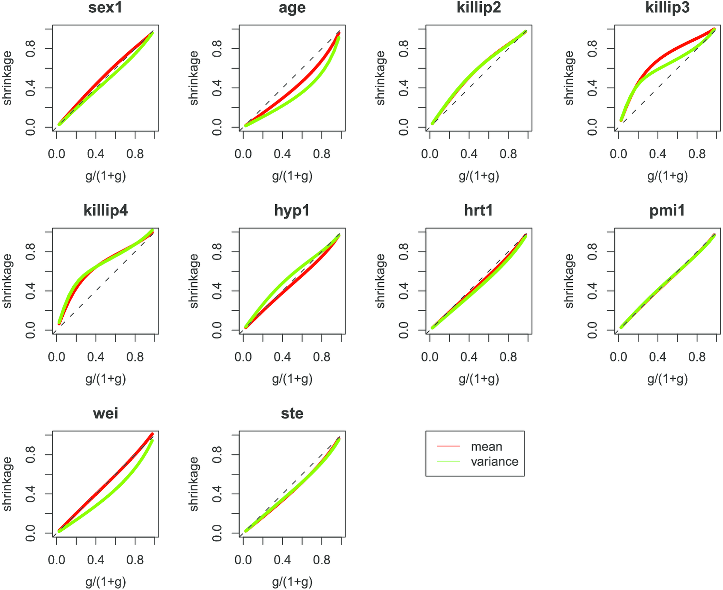}

\caption{Shrinkage of posterior means and variances of regression
coefficients under the generalized $g$-prior for various values of
$g$. The posterior distribution has been calculated with the {\tt
R-INLA} software and the empirical shrinkage is plotted against the
theoretical shrinkage $g/(g+1)$.}
\label{fig:gusto-shrinkage}
\end{figure*}

Figure~\ref{fig:gusto-shrinkage} shows that
there is a good agreement between empirical and
theoretical shrinkage $g/(g+1)$ for most regression coefficients, which supports
the validity of the approximation
\eqref{eq:approx-gaussian-posterior}. The agreement is not so good
for $x_{2}$ (Age) and the
factor variable $x_{3}$ (Killip class), perhaps because the
strong degree of discrimination of these important predictors may
affect the validity of the approximation $\fisher(\alpha,
\bolds{\beta}_{j}) \approx\fisher(\alpha,
\mathbf{0})$ from Section~\ref{sec:role-g-shrinkage}.

\subsection{Bootstrap Cross-Validation}
\label{sec:progn-modell-30:bootstrap-study}

To quantify and compare the predictive performance of the TBF methods,
we have performed a bootstrap cross-validation study. To reduce
computation time, we have used the best 8000 models based on a
stochastic model search, as described in \citet{sabanesbove.held2011a}
with 30,000
iterations, instead of exhaustive evaluation of all models. We have
used the area under the ROC curve (AUC, measures discrimination), the
calibration slope (CS) [\citet{cox1958}, measures calibration], and the
logarithmic score (LS) (measures both discrimination and calibration)
to quantify the predictive performance. See
\citet{GneitingRaftery2007} for a theoretical and
\citet{Steyerberg2009} for a more practical review of methods to
validate and compare probabilistic predictions. Both AUC and CS are 1
for perfect discrimination and calibration, respectively. In
practical applications they will be typically smaller than 1. The LS
is defined as $- \sum_{i=1}^{m} \log\{\hat{\pi}_{i}^{y_{i}}(1 -
\hat{\pi}_{i})^{1 - y_{i}}\}/m$, where $\hat{\pi}_{i}$ is the
predicted probability of death ($y_i=1$) for the $i$th patient in the
validation sample, $i=1,\ldots,m$. The LS is negatively oriented, that
is, the smaller, the better.

The apparent performance of the methods using the original sample
both for fitting and predicting is well-known to be of little value
for estimating the predictive performance for new data. Therefore, we
compute an estimate of the out-of-sample performance using bootstrap
cross-validation. For each of 1000 bootstrap samples, we fit the
methods and evaluate the above criteria based on the data not included
in the bootstrap sample.  We compare our methods with a more
traditional AIC- or BIC-based approach for (Bayesian) model selection
and averaging based on
posterior model probabilities proportional to $\exp(-\mathrm{AIC}_j/2)$
and $\exp(-\mathrm{BIC}_j/2)$, respectively [see
\citet{Claeskens.Hjort2008}], and to the \citet{hu.johnson2009} choice
$g = 2  n$. In addition, we apply a recently developed method for
variable selection in generalized additive models to our setting
[\citet{marra.wood2011}, Section~2.1]. The method gives component-wise
shrinkage of covariate effects included, similar to a Bayesian model
average (BMA). Finally, simple backward selection with AIC or BIC has
been included as well as just fitting the full model.

\begin{table}
\caption{GUSTO-I data: Comparison of the predictive performance of
variable selection using bootstrap
cross-validation of AUC, Calibration slope (CS), and Logarithmic score (LS)}
\label{tab:gusto-bootstrap-hyperg-results}
\begin{tabular*}{\columnwidth}{@{\extracolsep{\fill}}lcccc@{}}
\hline
& & \textbf{AUC} & \textbf{CS} & \textbf{LS} \\
\hline
Local EB & MAP & 0.8313 & 0.8643 & 0.1874 \\
& MPM & 0.8322 & 0.8616 & 0.1870 \\
& BMA & 0.8344 & 0.8864 & 0.1860 \\[3pt]
Hyper-$g$ & MAP & 0.8314 & 0.8141 & 0.1880 \\
& MPM & 0.8322 & 0.8196 & 0.1876 \\
& BMA & 0.8343 & 0.8406 & 0.1865 \\[3pt]
Hyper-$g/n$ & MAP & 0.8310 & 0.8558 & 0.1877 \\
& MPM & 0.8320 & 0.8547 & 0.1872 \\
& BMA & 0.8345 & 0.8818 & 0.1860 \\[3pt]
ZS adapted & MAP & 0.8296 & 0.8396 & 0.1887 \\
& MPM & 0.8300 & 0.8398 & 0.1885 \\
& BMA & 0.8343 & 0.8662 & 0.1866 \\[3pt]
AIC & MAP & 0.8316 & 0.8208 & 0.1886 \\
& MPM & 0.8318 & 0.8271 & 0.1884 \\
& BMA & 0.8339 & 0.8492 & 0.1873 \\[3pt]
BIC & MAP & 0.8259 & 0.8415 & 0.1908 \\
& MPM & 0.8261 & 0.8424 & 0.1907 \\
& BMA & 0.8313 & 0.8837 & 0.1884 \\[3pt]
Fixed $g=2n$ & MAP & 0.8250 & 0.8418 & 0.1906 \\
& MPM & 0.8251 & 0.8426 & 0.1905 \\
& BMA & 0.8308 & 0.8766 & 0.1881 \\[3pt]
GLM full & & 0.8314 & 0.8108 & 0.1888 \\
GLM select & & 0.8330 & 0.8787 & 0.1871 \\
Step AIC & &0.8314 & 0.8205 & 0.1887 \\
Step BIC & & 0.8285 & 0.8426 & 0.1898 \\
\hline
\end{tabular*}
\end{table}

The average criteria are shown in Table \ref{tab:gusto-bootstrap-hyperg-results}.
Considering first the
logarithmic score as our overall criterion, we see that, for any of
the four methods to estimate $g$ based on TBFs, BMA is better
than MPM, and MPM is better than MAP, and this is also true for
AUC. This is not surprising, given the theoretical advantage of BMA
over single models concerning prediction. The empirical superiority of
MPM over MAP indicates that the theoretical superiority of the MPM
approach in the linear model may extend to GLMs. We note that the BMA
is also superior in terms of calibration, whereas there is no clear
preference for either MAP or MPM in terms of CS. Overall, the local EB
approach performs best, closely followed by hyper-$g/n$. We would have
expected more similarities between local EB and \mbox{hyper-$g$}, which is
substantially worse, in particular, in terms of calibration. The ZS
adapted approach is better than hyper-$g$ in terms of calibration, but
slightly worse in terms of discrimination and LS.

Considering the alternatives to the TBF approach, AIC-weighted model
selection has a similar performance to hyper-$g$ and ZS adapted, but
is not as good as local EB or hyper-$g/n$. BIC-weighted model
selection and fixing $g$ at $2 n$ perform substantially worse, and so
do the two stepwise procedures. Simply using the full model gives
reasonable discrimination, but very poor calibration, and so the LS is
very poor. Among the alternative methods, the variable selection
according to \citet{marra.wood2011} (``GLM Select'') performs best.
Its additional flexibility from separate shrinkage of the coefficients
leads to a similar performance as our (global shrinkage) MPM model
with either local EB or hyper-$g/n$. However, it is not as good as the
BMAs (which also have implicit coefficient-wise shrinkage) with any of
our four approaches.

\section{Discussion}
\label{sec:discussion}
In this paper we considered test-based Bayes factors derived from the
deviance statistic for generalized linear
models, emphasizing
that the implicitly used prior on the regression coefficients is a generalized
$g$-prior. As with the data-based Bayes factors, estimation of $g$ is possible
and recommended. Local EB estimation of $g$ leads to posterior means
of the regression coefficients that correspond to shrinkage estimates
from the
literature. Alternatively, full Bayes estimation of $g$ is possible and
leads to
generalized hyper-$g$ priors.

In an empirical comparison, the TBFs have been shown to be in good
agreement with the corresponding DBFs. We developed a bias-correction
in the linear model under empirical Bayes which has further reduced
the error. It will be interesting to develop similar corrections for
the fully Bayesian approaches. Another important area of theoretical
research would be to investigate the conditions for optimality of the
MPM model in GLMs.

TBFs are applicable in a wider context. In particular, the proposed
methodology can be used for function selection
[\citet{sabanesbove.held2011a}] and can be
extended to the Cox proportional hazards model,
which we will report elsewhere. Also, regression models for
multicategorical data such as the proportional odds model or the
multinomial logistic regression model return a deviance, so the TBF
approach will be applicable in these settings. The same is true for
CART models [\citet{gravestock2014}] and mixed models with fixed (known)
random effects variances, where a (marginal) deviance is also
available. This is important in our context, as it would allow us to
combine the spline-based Bayesian model and function selection
[\citet{sabanesbove.etal2014}] with TBFs. However, more research
on the asymptotic distribution of the deviance is needed
for the application of TBFs to mixed models with unknown variance components.

\begin{appendix}

\section{Proofs for Section~\texorpdfstring{\protect\ref{SEC:TEST-BASED-BAYES:ASYMPT-DISTR-DEVI}}{2.2.1}}
\label{app:proof-test-based-bayes}

In Section~\ref{SEC:TEST-BASED-BAYES:ASYMPT-DISTR-DEVI} we state that the
distribution of the deviance converges for $n \to\infty$ to a noncentral
chi-squared distribution with $d_{j}$ degrees of freedom, where
$\lambda
_{j} =
\bolds{\beta}_{j}^{\top} \fisher_{\bolds{\beta}_{j},
\bolds{\beta}_{j}} \bolds{\beta}_{j}$ is the noncentrality
parameter. This is essentially proven by \citet{davidson.lever1970},
and we
briefly show how their Theorem~1 applies here. In their notation the
model is
parametrized by $\bolds{\theta} = (\bolds{\theta
}_{1}^\top,
\theta_{2})^\top$ with $\bolds{\theta}_{1} =
\bolds{\beta}_{j}$ being the parameter of interest and $\theta
_{2} =
\alpha$ being the nuisance parameter. We test the null hypothesis
$H_0$: $\bolds{\theta} = \bolds{\theta}_{0} =
(\mathbf
{0}^\top, \theta_{2})^\top$. We consider a sequence of
local alternatives $\bolds{\theta}^{n} =
(\bolds{\theta}_{1}^{n}, \theta_{2})$ with components
$\theta_{1k}^{n} = \delta_{k} / \sqrt{n}$ of
$\bolds{\theta}_{1}^{n}$, where $\delta_k \neq0$, $k=1,\ldots,
d_{j}$. It follows that $\bolds{\theta}^{n} \to
\bolds{\theta}_{0}$ for $n \to\infty$. Then Theorem~1
of \citet{davidson.lever1970} states that for $n \to\infty$ the
deviance converges in distribution to a noncentral chi-squared
distribution with $d_{j}$ degrees of freedom and noncentrality
parameter
$\bolds{\delta}^{\top}\overline{\mathbf
{C}}_{11}(\bolds
{\theta}_{0})\bolds{\delta}$, where
$\bolds{\delta} = (\delta_1,\ldots, \delta_{d_j})^{\top}$. Here
$\overline{\mathbf{C}}_{11}(\bolds{\theta}_{0})$ is the
inverse of the
submatrix corresponding to $\bolds{\theta}_{1}$ of the inverse
expected Fisher
information from one observation, evaluated at $\bolds{\theta} =
\bolds{\theta}_{0}$. But we know that the expected Fisher
information is
block-diagonal for $\bolds{\theta} = \bolds{\theta}_{0}$, so
$\overline{C}_{11}(\bolds{\theta}_{0})$ is just the submatrix of
the expected
Fisher information from one observation. Moreover, for
$n$ observations we have $\fisher_{\bolds{\beta}_{j},
\bolds{\beta}_{j}} = n \cdot\overline{\mathbf
{C}}_{11}(\bolds{\theta}_{0})$,
and combined with $\bolds{\delta} = \sqrt{n}
\bolds{\beta}_{j}$, we obtain the noncentrality parameter
$\lambda_{j} = \bolds{\beta}_{j}^{\top}
\fisher_{\bolds{\beta}_{j}, \bolds{\beta}_{j}}
\bolds{\beta}_{j}$.

In order to derive the prior distribution for $\lambda_{j}$ based on
the generalized
$g$-prior \eqref
{eq:g-priors-generalized-linear-model:fisher-representation} for
$\bolds{\beta}_{j}$ as stated in
Section~\ref{sec:test-based-bayes:defining-test-based}, first note
that the
generalized $g$-prior corresponds to
\[
\tilde{\bolds{\beta}}_{j} = \bigl( \fisher_{\bolds{\beta}_{j}, \bolds{\beta}_{j}}^{1/2}
/ \sqrt {g} \bigr) \bolds{\beta}_{j} \sim \Nor_{d_{j}}(
\mathbf{0}, \IdMat_{d_{j}}),
\]
where $\fisher_{\bolds{\beta}_{j}, \bolds{\beta}_{j}}^{1/2}$
is the
upper-triangular Cholesky root of $\fisher_{\bolds{\beta}_{j},
\bolds{\beta}_{j}}$. Hence,
$\tilde{\bolds{\beta}}^{\top}_{j}\tilde{\bolds{\beta}}_{j}
\sim
\chi^{2}(d_{j})$, which is a $\Ga(d_{j}/2,  1/2)$
distribution. Expanding the quadratic form, we obtain
\begin{eqnarray*}
\tilde{\bolds{\beta}}^{\top}_{j}\tilde{\bolds{\beta}}_{j} &=& 1/\sqrt{g} \bolds{\beta}_{j}^{\top}
\fisher_{\bolds
{\beta}_{j},
\bolds{\beta}_{j}}^{\top/2} \fisher_{\bolds{\beta}_{j},
\bolds{\beta}_{j}}^{1/2}
\bolds{\beta}_{j} 1/\sqrt{g}\\
 &=& 1/g \bolds{\beta}_{j}^{\top} \fisher_{\bolds{\beta}_{j},
  \bolds{\beta}_{j}} \bolds{\beta}_{j} = \lambda_{j} / g
\end{eqnarray*}
and, finally, $\lambda_{j} = g \cdot\lambda_{j} / g \sim\Ga
(d_{j}/2, 1/(2g))$.

\section{Proofs for Section~\texorpdfstring{\protect\ref{SEC:ESTIMATING-G-VIA-EB}}{3.2}}
\label{app:proof-estimating-g-via-EB}

For ease of notation we drop the index $j$ of the alternative model
and simply denote the deviance with $z$, the associated degrees of
freedom with $d$, while $\TBFs$ denotes the corresponding TBF with
respect to the null model.

For the bounds mentioned in Section~\ref{SEC:ESTIMATING-G-VIA-EB}
usually the
minimum Bayes factor in favor of the null hypothesis is considered,
which is
$\mTBFs^{-1}$ in our notation.
Let the $P$-value\vspace*{1pt} be $p = 1 - \mathrm{F}_{\chi^{2}(d)}(z)$, where
$\mathrm{F}_{\chi^{2}(d)}$ is the cumulative distribution\vspace*{1pt} function of the
chi-squared distribution with $d$ degrees of freedom.
The proofs are adapted from
\citet{malaguerra2012}:
\begin{longlist}[1.]
\item[1.] Let $d=1$ and $z > d = 1$.
Let $q = \Phi^{-1}(1 - p/2)$ be the corresponding
quantile of the standard normal distribution with
cumulative distribution function~$\Phi $.
We have $q^{2} = z$ since a
squared standard normal random variable is $\chi^{2}(1)$-distributed
and, hence,
$\mTBFs^{-1} =
z^{1/2}\exp(-z/2) \exp(1/2) =
q \exp(-q^{2}/2) \sqrt{e}$,
which is the required result from \citet{berger.sellke1987}.
\item[2.] Let $d=2$ and $z > d = 2$. Due to
$\mathrm{F}_{\chi^{2}(2)}(z) = 1 - \exp(- z/2)$, we have $p = \exp
(-z/2)$ or
$z=-2\log(p)$, such that $z > 2$ is equivalent to $p < 1/e$. Moreover,
$\mTBFs^{-1} =
(2/z)^{-1} \exp
 (
- ({z - 2})/{2}
 ) =
- e  p  \log(p)$,
which is the required result from \citet{sellke.etal2001}.
\item[3.] The universal bound from \citet{EdwardsLindmanSavage1963} that we
want to
reach is $\exp(-q^{2}/2)$, here $q = \Phi^{-1}(1 - p)$.
We have to show that for $d \to\infty$ and
fixed $P$-value, the ratio of $\mTBFs^{-1} $ and this universal
bound is 1. With $d \to\infty$ we have $(z - d)/\sqrt{2d} \sima
\Nor(0, 1)$ and, hence, 
$z \approx d + \sqrt{2d} q$.
Plugging this in \eqref{eq:maximum-tbf}, we obtain
\begin{eqnarray*}
&&\frac{\mTBFs^{-1}}{\exp(-q^{2}/2)} \\
&&\quad\approx\biggl( \frac{d}{\sqrt{2d}q + d} \biggr)^{-d/2}
\exp \biggl( - \sqrt{\frac{d}{2}}q +q^{2}/2 \biggr)
\\
&&\quad= \exp \bigl\{-aq + a^{2}\log(1 + q/a) + q^{2}/2 \bigr\}
\end{eqnarray*}
with $a = \sqrt{d/2}$. Now for large $d$ the term $q/a$ is
small and,
hence, we can apply a second-order Taylor expansion of $\log(1 + x)$
around $x=0$,
giving $\log(1 + x) \approx x - x^{2} / 2$, and we obtain
\begin{eqnarray*}
\frac{\mTBFs^{-1}}{\exp(-q^{2}/2)} &\approx& \exp \biggl\{-aq + a^{2} \biggl(
\frac{q}{a} - \frac{q^{2}}{2a^{2}} \biggr) + \frac{q^{2}}{2} \biggr\} \\
&=&
\exp(0) = 1,
\end{eqnarray*}
which proves the statement.
\end{longlist}
\end{appendix}
\section*{Acknowledgments}
We thank Kerry L. Lee and Ewout W. Steyerberg for permission to use
the GUSTO-I data set. We are also grateful to Rafael Sauter for help
with the \texttt{R-INLA} software in Section~\ref{sec:progn-modell-30:shrinkage}
and to Manuela Ott for
proofreading the final manuscript. We finally acknowledge helpful
comments by
two referees on an earlier version of this article.


\begin{thebibliography}{41}

\bibitem[\protect\citeauthoryear{Barbieri and
Berger}{2004}]{BarbieriBerger2004}
%
\begin{barticle}[mr]
\bauthor{\bsnm{Barbieri},~\bfnm{Maria~Maddalena}\binits{M.~M.}}
\AND
\bauthor{\bsnm{Berger},~\bfnm{James~O.}\binits{J.~O.}}
(\byear{2004}).
\btitle{Optimal predictive model selection}.
\bjournal{Ann. Statist.}
\bvolume{32}
\bpages{870--897}.
\bid{doi={10.1214/009053604000000238}, issn={0090-5364}, mr={2065192}}
\end{barticle}
%

\bptok{imsref}%
\endbibitem

\bibitem[\protect\citeauthoryear{Bayarri et~al.}{2012}]{bayarri.etal2012}
%
\begin{barticle}[mr]
\bauthor{\bsnm{Bayarri},~\bfnm{M.~J.}\binits{M.~J.}},
\bauthor{\bsnm{Berger},~\bfnm{J.~O.}\binits{J.~O.}},
\bauthor{\bsnm{Forte},~\bfnm{A.}\binits{A.}} \AND
\bauthor{\bsnm{Garc{\'{\i}}a-Donato},~\bfnm{G.}\binits{G.}}
(\byear{2012}).
\btitle{Criteria for {B}ayesian model choice with application to
variable selection}.
\bjournal{Ann. Statist.}
\bvolume{40}
\bpages{1550--1577}.
\bid{doi={10.1214/12-AOS1013}, issn={0090-5364}, mr={3015035}}
\end{barticle}
%

\bptok{imsref}%
\endbibitem

\bibitem[\protect\citeauthoryear{Berger and
Pericchi}{2001}]{BergerPericchi2001}
%
\begin{bincollection}[mr]
\bauthor{\bsnm{Berger},~\bfnm{James~O.}\binits{J.~O.}} \AND
\bauthor{\bsnm{Pericchi},~\bfnm{Luis~R.}\binits{L.~R.}}
(\byear{2001}).
\btitle{Objective {B}ayesian methods for model selection: Introduction
and comparison}.
In \bbooktitle{Model Selection}
(\beditor{\binits{P.}\bfnm{P.} \bsnm{Lahiri}}, eds.).
\bseries{Institute of Mathematical Statistics Lecture
Notes---Monograph Series}
\bvolume{38}
\bpages{135--207}.
\bpublisher{IMS},
\blocation{Beachwood, OH}.
\bid{doi={10.1214/lnms/1215540968}, mr={2000753}}
\bptnote{check related}%
\end{bincollection}
%

\bptok{imsref}%
\endbibitem

\bibitem[\protect\citeauthoryear{Berger and Sellke}{1987}]{berger.sellke1987}
%
\begin{barticle}[mr]
\bauthor{\bsnm{Berger},~\bfnm{James~O.}\binits{J.~O.}} \AND
\bauthor{\bsnm{Sellke},~\bfnm{Thomas}\binits{T.}}
(\byear{1987}).
\btitle{Testing a point null hypothesis: Irreconcilability of
{$p$}-values and evidence}.
\bjournal{J. Amer. Statist. Assoc.}
\bvolume{82}
\bpages{112--139}.
\bid{issn={0162-1459}, mr={0883340}}
\bptnote{check related}%
\end{barticle}
%

\bptok{imsref}%
\endbibitem

\bibitem[\protect\citeauthoryear{Claeskens and
Hjort}{2008}]{Claeskens.Hjort2008}
%
\begin{bbook}[mr]
\bauthor{\bsnm{Claeskens},~\bfnm{Gerda}\binits{G.}} \AND
\bauthor{\bsnm{Hjort},~\bfnm{Nils~Lid}\binits{N.~L.}}
(\byear{2008}).
\btitle{Model Selection and Model Averaging}.
\bpublisher{Cambridge Univ. Press},
\blocation{Cambridge}.
\bid{doi={10.1017/CBO9780511790485}, mr={2431297}}
\end{bbook}
%

\bptok{imsref}%
\endbibitem

\bibitem[\protect\citeauthoryear{Copas}{1983}]{copas1983}
%
\begin{barticle}[mr]
\bauthor{\bsnm{Copas},~\bfnm{J.~B.}\binits{J.~B.}}
(\byear{1983}).
\btitle{Regression, prediction and shrinkage}.
\bjournal{J. R. Stat. Soc. Ser. B. Stat. Methodol.}
\bvolume{45}
\bpages{311--354}.
\bid{issn={0035-9246}, mr={0737642}}
\end{barticle}
%

\bptok{imsref}%
\endbibitem

\bibitem[\protect\citeauthoryear{Copas}{1997}]{Copas1997}
%
\begin{barticle}[auto:parserefs-M02]
\bauthor{\bsnm{Copas},~\bfnm{J.~B.}\binits{J.~B.}}
(\byear{1997}).
\btitle{{U}sing regression models for prediction: Shrinkage and
regression to the mean}.
\bjournal{Stat. Methods Med. Res.}
\bvolume{6}
\bpages{167--183}.
\end{barticle}
%

\bptok{imsref}%
\endbibitem

\bibitem[\protect\citeauthoryear{Cox}{1958}]{cox1958}
%
\begin{barticle}[auto:parserefs-M02]
\bauthor{\bsnm{Cox},~\bfnm{D.~R.}\binits{D.~R.}}
(\byear{1958}).
\btitle{{T}wo further applications of a model for binary regression}.
\bjournal{Biometrika}
\bvolume{45}
\bpages{562--565}.
\end{barticle}
%

\bptok{imsref}%
\endbibitem

\bibitem[\protect\citeauthoryear{Cui and George}{2008}]{CuiGeorge2008}
%
\begin{barticle}[mr]
\bauthor{\bsnm{Cui},~\bfnm{Wen}\binits{W.}} \AND
\bauthor{\bsnm{George},~\bfnm{Edward~I.}\binits{E.~I.}}
(\byear{2008}).
\btitle{Empirical {B}ayes vs. fully {B}ayes variable selection}.
\bjournal{J. Statist. Plann. Inference}
\bvolume{138}
\bpages{888--900}.
\bid{doi={10.1016/j.jspi.2007.02.011}, issn={0378-3758}, mr={2416869}}
\end{barticle}
%

\bptok{imsref}%
\endbibitem

\bibitem[\protect\citeauthoryear{Davidson and
Lever}{1970}]{davidson.lever1970}
%
\begin{barticle}[mr]
\bauthor{\bsnm{Davidson},~\bfnm{Roger~R.}\binits{R.~R.}} \AND
\bauthor{\bsnm{Lever},~\bfnm{William~E.}\binits{W.~E.}}
(\byear{1970}).
\btitle{The limiting distribution of the likelihood ratio statistic
under a class of local alternatives}.
\bjournal{Sankhy\=a Ser. A}
\bvolume{32}
\bpages{209--224}.
\bid{issn={0581-572X}, mr={0297050}}
\end{barticle}
%

\bptok{imsref}%
\endbibitem

\bibitem[\protect\citeauthoryear{Edwards, Lindman and
Savage}{1963}]{EdwardsLindmanSavage1963}
%
\begin{barticle}[auto:parserefs-M02]
\bauthor{\bsnm{Edwards},~\bfnm{W.}\binits{W.}},
\bauthor{\bsnm{Lindman},~\bfnm{H.}\binits{H.}} \AND
\bauthor{\bsnm{Savage},~\bfnm{L.~J.}\binits{L.~J.}}
(\byear{1963}).
\btitle{{B}ayesian statistical inference for psychological research}.
\bjournal{Psychological Review}
\bvolume{70}
\bpages{193--242}.
\end{barticle}
%

\bptok{imsref}%
\endbibitem

\bibitem[\protect\citeauthoryear{Fern{\'a}ndez, Ley and
Steel}{2001}]{FernandezLeySteel2001}
%
\begin{barticle}[mr]
\bauthor{\bsnm{Fern{\'a}ndez},~\bfnm{Carmen}\binits{C.}},
\bauthor{\bsnm{Ley},~\bfnm{Eduardo}\binits{E.}} \AND
\bauthor{\bsnm{Steel},~\bfnm{Mark~F.~J.}\binits{M.~F.~J.}}
(\byear{2001}).
\btitle{Benchmark priors for {B}ayesian model averaging}.
\bjournal{J. Econometrics}
\bvolume{100}
\bpages{381--427}.
\bid{doi={10.1016/S0304-4076(00)00076-2}, issn={0304-4076}, mr={1820410}}
\end{barticle}
%

\bptok{imsref}%
\endbibitem

\bibitem[\protect\citeauthoryear{Foster and George}{1994}]{foster.george1994}
%
\begin{barticle}[mr]
\bauthor{\bsnm{Foster},~\bfnm{Dean~P.}\binits{D.~P.}} \AND
\bauthor{\bsnm{George},~\bfnm{Edward~I.}\binits{E.~I.}}
(\byear{1994}).
\btitle{The risk inflation criterion for multiple regression}.
\bjournal{Ann. Statist.}
\bvolume{22}
\bpages{1947--1975}.
\bid{doi={10.1214/aos/1176325766}, issn={0090-5364}, mr={1329177}}
\end{barticle}
%

\bptok{imsref}%
\endbibitem

\bibitem[\protect\citeauthoryear{Geisser}{1984}]{geisser1984}
%
\begin{barticle}[mr]
\bauthor{\bsnm{Geisser},~\bfnm{Seymour}\binits{S.}}
(\byear{1984}).
\btitle{On prior distributions for binary trials}.
\bjournal{Amer. Statist.}
\bvolume{38}
\bpages{244--251}.
\bid{doi={10.2307/2683393}, issn={0003-1305}, mr={0770258}}
\bptnote{check related}%
\end{barticle}
%

\bptok{imsref}%
\endbibitem

\bibitem[\protect\citeauthoryear{George and Foster}{2000}]{GeorgeFoster2000}
%
\begin{barticle}[mr]
\bauthor{\bsnm{George},~\bfnm{Edward~I.}\binits{E.~I.}} \AND
\bauthor{\bsnm{Foster},~\bfnm{Dean~P.}\binits{D.~P.}}
(\byear{2000}).
\btitle{Calibration and empirical {B}ayes variable selection}.
\bjournal{Biometrika}
\bvolume{87}
\bpages{731--747}.
\bid{doi={10.1093/biomet/87.4.731}, issn={0006-3444}, mr={1813972}}
\end{barticle}
%

\bptok{imsref}%
\endbibitem

\bibitem[\protect\citeauthoryear{Gneiting and
Raftery}{2007}]{GneitingRaftery2007}
%
\begin{barticle}[mr]
\bauthor{\bsnm{Gneiting},~\bfnm{Tilmann}\binits{T.}} \AND
\bauthor{\bsnm{Raftery},~\bfnm{Adrian~E.}\binits{A.~E.}}
(\byear{2007}).
\btitle{Strictly proper scoring rules, prediction, and estimation}.
\bjournal{J. Amer. Statist. Assoc.}
\bvolume{102}
\bpages{359--378}.
\bid{doi={10.1198/016214506000001437}, issn={0162-1459}, mr={2345548}}
\end{barticle}
%

\bptok{imsref}%
\endbibitem

\bibitem[\protect\citeauthoryear{Goodman}{1999a}]{goodman19991}
%
\begin{barticle}[auto:parserefs-M02]
\bauthor{\bsnm{Goodman},~\bfnm{S.~N.}\binits{S.~N.}}
(\byear{1999}a).
\btitle{{T}oward evidence-based medical statistics. 1: {T}he
{$P$}-value fallacy}.
\bjournal{Annals of Internal Medicine}
\bvolume{130}
\bpages{995--1004}.
\end{barticle}
%

\bptok{imsref}%
\endbibitem

\bibitem[\protect\citeauthoryear{Goodman}{1999b}]{goodman19992}
%
\begin{barticle}[auto:parserefs-M02]
\bauthor{\bsnm{Goodman},~\bfnm{S.~N.}\binits{S.~N.}}
(\byear{1999}b).
\btitle{{T}oward evidence-based medical statistics. 2: {T}he {B}ayes factor}.
\bjournal{Annals of Internal Medicine}
\bvolume{130}
\bpages{1005--1013}.
\end{barticle}
%

\bptok{imsref}%
\endbibitem

\bibitem[\protect\citeauthoryear{Gravestock}{2014}]{gravestock2014}
%
\begin{bmisc}[auto:parserefs-M02]
\bauthor{\bsnm{Gravestock},~\bfnm{I.}\binits{I.}}
(\byear{2014}).
\bhowpublished{{B}ayesian tree models priors and posterior approximations.
Master's thesis,
Univ. Zurich}.
\end{bmisc}
%

\bptok{imsref}%
\endbibitem

\bibitem[\protect\citeauthoryear{Held}{2010}]{held2010}
%
\begin{barticle}[auto:parserefs-M02]
\bauthor{\bsnm{Held},~\bfnm{L.}\binits{L.}}
(\byear{2010}).
\btitle{{A} nomogram for $P$-values}.
\bjournal{BMC Medical Research Methodology}
\bvolume{10}
\bpages{21}.
\end{barticle}
%

\bptok{imsref}%
\endbibitem

\bibitem[\protect\citeauthoryear{Hjort and
Claeskens}{2003}]{HjortClaeskens2003}
%
\begin{barticle}[mr]
\bauthor{\bsnm{Hjort},~\bfnm{Nils~Lid}\binits{N.~L.}} \AND
\bauthor{\bsnm{Claeskens},~\bfnm{Gerda}\binits{G.}}
(\byear{2003}).
\btitle{Frequentist model average estimators}.
\bjournal{J. Amer. Statist. Assoc.}
\bvolume{98}
\bpages{879--899}.
\bid{doi={10.1198/016214503000000828}, issn={0162-1459}, mr={2041481}}
\end{barticle}
%

\bptok{imsref}%
\endbibitem

\bibitem[\protect\citeauthoryear{Hu and Johnson}{2009}]{hu.johnson2009}
%
\begin{barticle}[mr]
\bauthor{\bsnm{Hu},~\bfnm{Jianhua}\binits{J.}} \AND
\bauthor{\bsnm{Johnson},~\bfnm{Valen~E.}\binits{V.~E.}}
(\byear{2009}).
\btitle{Bayesian model selection using test statistics}.
\bjournal{J. R. Stat. Soc. Ser. B. Stat. Methodol.}
\bvolume{71}
\bpages{143--158}.
\bid{doi={10.1111/j.1467-9868.2008.00678.x}, issn={1369-7412}, mr={2655527}}
\end{barticle}
%

\bptok{imsref}%
\endbibitem

\bibitem[\protect\citeauthoryear{Johnson}{2005}]{johnson2005}
%
\begin{barticle}[mr]
\bauthor{\bsnm{Johnson},~\bfnm{Valen~E.}\binits{V.~E.}}
(\byear{2005}).
\btitle{Bayes factors based on test statistics}.
\bjournal{J. R. Stat. Soc. Ser. B. Stat. Methodol.}
\bvolume{67}
\bpages{689--701}.
\bid{doi={10.1111/j.1467-9868.2005.00521.x}, issn={1369-7412}, mr={2210687}}
\end{barticle}
%

\bptok{imsref}%
\endbibitem

\bibitem[\protect\citeauthoryear{Johnson}{2008}]{johnson2008}
%
\begin{barticle}[mr]
\bauthor{\bsnm{Johnson},~\bfnm{Valen~E.}\binits{V.~E.}}
(\byear{2008}).
\btitle{Properties of {B}ayes factors based on test statistics}.
\bjournal{Scand. J. Stat.}
\bvolume{35}
\bpages{354--368}.
\bid{doi={10.1111/j.1467-9469.2007.00576.x}, issn={0303-6898}, mr={2418746}}
\end{barticle}
%

\bptok{imsref}%
\endbibitem

\bibitem[\protect\citeauthoryear{Kass and
Wasserman}{1995}]{KassWasserman1995}
%
\begin{barticle}[mr]
\bauthor{\bsnm{Kass},~\bfnm{Robert~E.}\binits{R.~E.}} \AND
\bauthor{\bsnm{Wasserman},~\bfnm{Larry}\binits{L.}}
(\byear{1995}).
\btitle{A reference {B}ayesian test for nested hypotheses and its
relationship to the {S}chwarz criterion}.
\bjournal{J. Amer. Statist. Assoc.}
\bvolume{90}
\bpages{928--934}.
\bid{issn={0162-1459}, mr={1354008}}
\end{barticle}
%

\bptok{imsref}%
\endbibitem

\bibitem[\protect\citeauthoryear{Lawless and
Singhal}{1978}]{lawless.singhal1978}
%
\begin{barticle}[auto:parserefs-M02]
\bauthor{\bsnm{Lawless},~\bfnm{J.~F.}\binits{J.~F.}} \AND
\bauthor{\bsnm{Singhal},~\bfnm{K.}\binits{K.}}
(\byear{1978}).
\btitle{{E}fficient screening of nonnormal regression models}.
\bjournal{Biometrics}
\bvolume{34}
\bpages{318--327}.
\end{barticle}
%

\bptok{imsref}%
\endbibitem

\bibitem[\protect\citeauthoryear{Lee et~al.}{1995}]{lee.etal1995}
%
\begin{barticle}[auto:parserefs-M02]
\bauthor{\bsnm{Lee},~\bfnm{K.~L.}\binits{K.~L.}},
\bauthor{\bsnm{Woodlief},~\bfnm{L.~H.}\binits{L.~H.}},
\bauthor{\bsnm{Topol},~\bfnm{E.~J.}\binits{E.~J.}},
\bauthor{\bsnm{Weaver},~\bfnm{W.~D.}\binits{W.~D.}},
\bauthor{\bsnm{Betriu},~\bfnm{A.}\binits{A.}},
\bauthor{\bsnm{Col},~\bfnm{J.}\binits{J.}},
\bauthor{\bsnm{Simoons},~\bfnm{M.}\binits{M.}},
\bauthor{\bsnm{Aylward},~\bfnm{P.}\binits{P.}},
\bauthor{\bsnm{Van~de Werf},~\bfnm{F.}\binits{F.}} \AND
\bauthor{\bsnm{Califf},~\bfnm{R.~M.}\binits{R.~M.}}
(\byear{1995}).
\btitle{{P}redictors of 30-day mortality in the era of reperfusion for
acute myocardial infarction: Results from an international trial of
41,021 patients}.
\bjournal{Circulation}
\bvolume{91}
\bpages{1659--1668}.
\end{barticle}
%

\bptok{imsref}%
\endbibitem

\bibitem[\protect\citeauthoryear{Liang
et~al.}{2008}]{LiangPauloMolinaClydeBerger2008}
%
\begin{barticle}[mr]
\bauthor{\bsnm{Liang},~\bfnm{Feng}\binits{F.}},
\bauthor{\bsnm{Paulo},~\bfnm{Rui}\binits{R.}},
\bauthor{\bsnm{Molina},~\bfnm{German}\binits{G.}},
\bauthor{\bsnm{Clyde},~\bfnm{Merlise~A.}\binits{M.~A.}} \AND
\bauthor{\bsnm{Berger},~\bfnm{Jim~O.}\binits{J.~O.}}
(\byear{2008}).
\btitle{Mixtures of {$g$} priors for {B}ayesian variable selection}.
\bjournal{J. Amer. Statist. Assoc.}
\bvolume{103}
\bpages{410--423}.
\bid{doi={10.1198/016214507000001337}, issn={0162-1459}, mr={2420243}}
\end{barticle}
%

\bptok{imsref}%
\endbibitem

\bibitem[\protect\citeauthoryear{Lindley}{1957}]{Lindley1957}
%
\begin{barticle}[mr]
\bauthor{\bsnm{Lindley},~\bfnm{D.~V.}\binits{D.~V.}}
(\byear{1957}).
\btitle{{A} statistical paradox}.
\bjournal{Biometrika}
\bvolume{44}
\bpages{187--192}.
\end{barticle}
%

\bptok{imsref}%
\endbibitem

\bibitem[\protect\citeauthoryear{Malaguerra}{2012}]{malaguerra2012}
%
\begin{bmisc}[auto:parserefs-M02]
\bauthor{\bsnm{Malaguerra},~\bfnm{A.}\binits{A.}}
(\byear{2012}).
\bhowpublished{{B}ayesian variable selection based on test statistics.
Master's thesis,
Univ. Zurich}.
\end{bmisc}
%

\bptok{imsref}%
\endbibitem

\bibitem[\protect\citeauthoryear{Marra and Wood}{2011}]{marra.wood2011}
%
\begin{barticle}[mr]
\bauthor{\bsnm{Marra},~\bfnm{Giampiero}\binits{G.}} \AND
\bauthor{\bsnm{Wood},~\bfnm{Simon~N.}\binits{S.~N.}}
(\byear{2011}).
\btitle{Practical variable selection for generalized additive models}.
\bjournal{Comput. Statist. Data Anal.}
\bvolume{55}
\bpages{2372--2387}.
\bid{doi={10.1016/j.csda.2011.02.004}, issn={0167-9473}, mr={2786996}}
\end{barticle}
%

\bptok{imsref}%
\endbibitem

\bibitem[\protect\citeauthoryear{Nelder and
Wedderburn}{1972}]{NelderWedderburn1972}
%
\begin{barticle}[auto:parserefs-M02]
\bauthor{\bsnm{Nelder},~\bfnm{J.~A.}\binits{J.~A.}} \AND
\bauthor{\bsnm{Wedderburn},~\bfnm{R.~W.~M.}\binits{R.~W.~M.}}
(\byear{1972}).
\btitle{{G}eneralized linear models}.
\bjournal{J. Roy. Statist. Soc. Ser. A}
\bvolume{135}
\bpages{370--384}.
\end{barticle}
%

\bptok{imsref}%
\endbibitem


\bibitem[\protect\citeauthoryear{Rue, Martino and Chopin}{2009}]{naujas}
%
\begin{barticle}[auto:parserefs-M02]
\bauthor{\bsnm{Rue},~\bfnm{H.}\binits{H.}},
\bauthor{\bsnm{Martino},~\bfnm{S.}\binits{S.}} \AND
\bauthor{\bsnm{Chopin},~\bfnm{N.}\binits{N.}}
(\byear{2009}).
\btitle{Approximate Bayesian inference for
latent Gaussian models using integrated nested Laplace approximations
(with discussion)}.
\bjournal{J. Roy. Statist. Soc. Ser. B}
\bvolume{71}
\bpages{319--392}.
\end{barticle}
%

\bptok{imsref}%
\endbibitem


\bibitem[\protect\citeauthoryear{Saban{\'e}s~Bov{\'e} and
Held}{2011a}]{sabanesbove.held2011}
%
\begin{barticle}[mr]
\bauthor{\bsnm{Saban{\'e}s Bov{\'e}},~\bfnm{Daniel}\binits{D.}}
\AND
\bauthor{\bsnm{Held},~\bfnm{Leonhard}\binits{L.}}
(\byear{2011}a).
\btitle{Hyper-{$g$} priors for generalized linear models}.
\bjournal{Bayesian Anal.}
\bvolume{6}
\bpages{387--410}.
\bid{doi={10.1214/ba/1339616469}, issn={1936-0975}, mr={2843537}}
\end{barticle}
%

\bptok{imsref}%
\endbibitem

\bibitem[\protect\citeauthoryear{Saban{\'e}s~Bov{\'e} and
Held}{2011b}]{sabanesbove.held2011a}
%
\begin{barticle}[mr]
\bauthor{\bsnm{Saban{\'e}s Bov{\'e}},~\bfnm{Daniel}\binits{D.}}
\AND
\bauthor{\bsnm{Held},~\bfnm{Leonhard}\binits{L.}}
(\byear{2011}b).
\btitle{Bayesian fractional polynomials}.
\bjournal{Stat. Comput.}
\bvolume{21}
\bpages{309--324}.
\bid{doi={10.1007/s11222-010-9170-7}, issn={0960-3174}, mr={2806611}}
\end{barticle}
%

\bptok{imsref}%
\endbibitem

\bibitem[\protect\citeauthoryear{Saban{\'{e}}s~Bov{\'{e}}, Held and
Kauermann}{2014}]{sabanesbove.etal2014}
%
\begin{bmisc}[auto:parserefs-M02]
\bauthor{\bsnm{Saban{\'{e}}s Bov{\'{e}}},~\bfnm{D.}\binits{D.}},
\bauthor{\bsnm{Held},~\bfnm{L.}\binits{L.}} \AND
\bauthor{\bsnm{Kauermann},~\bfnm{G.}\binits{G.}}
(\byear{2014}).
\bhowpublished{{M}ixtures of $g$-priors for generalised additive model
selection with penalised splines.
\textit{J. Comput. Graph. Statist.}
DOI: \doiurl{10.1080/10618600.2014.912136}}.
\end{bmisc}
%

\bptok{imsref}%
\endbibitem

\bibitem[\protect\citeauthoryear{Scott and Berger}{2010}]{scott.berger2010}
%
\begin{barticle}[mr]
\bauthor{\bsnm{Scott},~\bfnm{James~G.}\binits{J.~G.}} \AND
\bauthor{\bsnm{Berger},~\bfnm{James~O.}\binits{J.~O.}}
(\byear{2010}).
\btitle{Bayes and empirical-{B}ayes multiplicity adjustment in the
variable-selection problem}.
\bjournal{Ann. Statist.}
\bvolume{38}
\bpages{2587--2619}.
\bid{doi={10.1214/10-AOS792}, issn={0090-5364}, mr={2722450}}
\end{barticle}
%

\bptok{imsref}%
\endbibitem

\bibitem[\protect\citeauthoryear{Sellke, Bayarri and
Berger}{2001}]{sellke.etal2001}
%
\begin{barticle}[mr]
\bauthor{\bsnm{Sellke},~\bfnm{Thomas}\binits{T.}},
\bauthor{\bsnm{Bayarri},~\bfnm{M.~J.}\binits{M.~J.}} \AND
\bauthor{\bsnm{Berger},~\bfnm{James~O.}\binits{J.~O.}}
(\byear{2001}).
\btitle{Calibration of {$p$}-values for testing precise null hypotheses}.
\bjournal{Amer. Statist.}
\bvolume{55}
\bpages{62--71}.
\bid{doi={10.1198/000313001300339950}, issn={0003-1305}, mr={1818723}}
\end{barticle}
%

\bptok{imsref}%
\endbibitem

\bibitem[\protect\citeauthoryear{Steyerberg}{2009}]{Steyerberg2009}
%
\begin{bbook}[auto:parserefs-M02]
\bauthor{\bsnm{Steyerberg},~\bfnm{E.}\binits{E.}}
(\byear{2009}).
\btitle{{C}linical {P}rediction {M}odels}.
\bpublisher{Springer},
\blocation{New York}.
\end{bbook}
%

\bptok{imsref}%
\endbibitem

\bibitem[\protect\citeauthoryear{van Houwelingen and
Le~Cessie}{1990}]{vanhouwelingen.lecessie1990}
%
\begin{barticle}[auto:parserefs-M02]
\bauthor{\bparticle{van} \bsnm{Houwelingen},~\bfnm{J.~C.}\binits
{J.~C.}} \AND
\bauthor{\bsnm{Le Cessie},~\bfnm{S.}\binits{S.}}
(\byear{1990}).
\btitle{{P}redictive value of statistical models}.
\bjournal{Stat. Med.}
\bvolume{9}
\bpages{1303--1325}.
\end{barticle}
%

\bptok{imsref}%
\endbibitem

\bibitem[\protect\citeauthoryear{Zellner}{1986}]{Zellner1986}
%
\begin{bincollection}[mr]
\bauthor{\bsnm{Zellner},~\bfnm{Arnold}\binits{A.}}
(\byear{1986}).
\btitle{On assessing prior distributions and {B}ayesian regression
analysis with {$g$}-prior distributions}.
In \bbooktitle{Bayesian Inference and Decision Techniques}
(\beditor{\binits{P.~K.}\bfnm{P.~K.} \bsnm{Goel}}
\AND
\beditor{\binits{A.}\bfnm{A.}~\bsnm{Zellner}}, eds.).
\bseries{Stud. Bayesian Econometrics Statist.}
\bvolume{6}
\bpages{233--243}.
\bpublisher{North-Holland},
\blocation{Amsterdam}.
\bid{mr={0881437}}
\end{bincollection}
%

\bptok{imsref}%
\endbibitem

\bibitem[\protect\citeauthoryear{Zellner and Siow}{1980}]{ZellnerSiow1980}
%
\begin{binproceedings}[auto:parserefs-M02]
\bauthor{\bsnm{Zellner},~\bfnm{A.}\binits{A.}} \AND
\bauthor{\bsnm{Siow},~\bfnm{A.}\binits{A.}}
(\byear{1980}).
\btitle{{P}osterior odds ratios for selected regression hypotheses}.
In \bbooktitle{Bayesian Statistics: Proceedings of the First
International Meeting Held in Valencia}
(\beditor{\bfnm{J.~M.}\binits{J.~M.}~\bsnm{Bernardo}},
\beditor{\bfnm{M.~H.}\binits{M.~H.}~\bsnm{DeGroot}},
\beditor{\bfnm{D.~V.}\binits{D.~V.}~\bsnm{Lindley}} \AND
\beditor{\bfnm{A.~F.~M.}\binits{A.~F.~M.}~\bsnm{Smith}}, eds.)
\bpages{585--603}.
\bpublisher{Univ. Valencia Press},
\blocation{Valencia}.
\end{binproceedings}
%
\bptok{imsref}%
\endbibitem
\end{thebibliography}
%
%




\end{document}